\pdfoutput=1

\documentclass[11pt,twoside,a4paper,cmspaper,final,collab]{cms-tdr}
\def\svnVersion{365774}\def\cmsCernNoTag{CERN-EP/2016-044}\def\cmsCernDate{\today}\def\cmsMessage{Published in the Journal of High Energy Physics as \href{http://dx.doi.org/10.1007/JHEP08(2016)029}{\doi{10.1007/JHEP08(2016)029.}}}

\begin{document}\cmsNoteHeader{TOP-13-004}

\hyphenation{had-ron-i-za-tion}
\hyphenation{cal-or-i-me-ter}
\hyphenation{de-vices}
\RCS$Revision: 364864 $
\RCS$HeadURL: svn+ssh://alverson@svn.cern.ch/reps/tdr2/papers/TOP-13-004/trunk/TOP-13-004.tex $
\RCS$Id: TOP-13-004.tex 364864 2016-08-18 10:17:11Z cdiez $
\newlength\cmsFigWidth
\ifthenelse{\boolean{cms@external}}{\setlength\cmsFigWidth{0.85\columnwidth}}{\setlength\cmsFigWidth{0.4\textwidth}}
\ifthenelse{\boolean{cms@external}}{\providecommand{\cmsLeft}{top}}{\providecommand{\cmsLeft}{left}}
\ifthenelse{\boolean{cms@external}}{\providecommand{\cmsRight}{bottom}}{\providecommand{\cmsRight}{right}}

\newcommand{\PV}{\HepParticle{V}{}{}\Xspace}
\newcommand{\Wjets}{\ensuremath{\PW+\text{jets}}\xspace}
\newcommand{\ttbg}{\ensuremath{\ttbar \text{ bkg}}\xspace}
 \newcommand{\eepm}{\ensuremath{\Pep\Pem}}
\newcommand{\mmpm}{\ensuremath{\Pgmp \Pgmm}}
\providecommand{\POWHEG} {\textsc{Powheg}\xspace}
\providecommand{\PYTHIA} {\textsc{Pythia}\xspace}
\providecommand{\HERWIG} {\textsc{Herwig}\xspace}
\newcommand{\stt}{\ensuremath{\sigma_{\ttbar}}}
\newcommand{\emu}{\ensuremath{\Pe\PGm}\xspace}
\newcommand{\Madspin} {\textsc{MadSpin}\xspace}
\newcommand{\ttbarV}{\ensuremath{\ttbar\PV}\xspace}

\cmsNoteHeader{TOP-13-004}
\title{Measurement of the $\ttbar$ production cross section in the $\Pe \mu$ channel in proton-proton collisions at $\sqrt{s} = 7$ and $8\TeV$}

\date{\today}

\abstract{
The inclusive cross section for top quark pair production is measured in proton-proton collisions at $\sqrt{s} = 7$ and $8\TeV$, corresponding to 5.0 and 19.7\fbinv, respectively, with the CMS experiment at the LHC. The cross sections are measured in the electron-muon channel using a binned likelihood fit to
multi-differential final state distributions related to identified \PQb quark jets and other jets in the event. The measured cross section values are $173.6 \pm 2.1\stat ^{+\,4.5}_{-\,4.0}\syst \pm 3.8\lum\unit{pb}$ at $\sqrt{s}=7\TeV$, and $244.9 \pm 1.4 \stat ^{+\,6.3}_{-\,5.5}\syst \pm 6.4\lum\unit{pb}$ at $\sqrt{s}=8\TeV$, in good agreement with QCD calculations at next-to-next-to-leading-order accuracy. The ratio of the cross sections measured at 7 and 8\TeV is determined, as well as cross sections in the fiducial regions defined by the acceptance requirements on the two charged leptons in the final state. The cross section results are used to determine the top quark pole mass via the dependence of the theoretically predicted cross section on the mass, giving a best result of $173.8^{+1.7}_{-1.8}\GeV$. The data at $\sqrt{s}=8\TeV$ are also used to set limits, for two neutralino mass values, on the pair production of supersymmetric partners of the top quark with masses close to the top quark mass.
}

\hypersetup{%
pdfauthor={CMS Collaboration},%
pdftitle={Measurement of the ttbar production cross section in the e mu
channel in proton-proton collisions at 7 and 8 TeV},%
pdfsubject={CMS},%
pdfkeywords={CMS, physics, software, computing}}

\maketitle
\section{Introduction }
\label{sec:intro}
The study of top quark pair (\ttbar) production in proton-proton (pp) collisions at the CERN LHC provides an important test of the standard model (SM). The total production cross section, $\sigma_{\ttbar}$,
can be accurately predicted
by quantum chromodynamics (QCD) calculations at next-to-next-to-leading order (NNLO).
A measurement of $\sigma_{\ttbar}$\/ can thus provide constraints on
essential ingredients in the calculation, such as the top quark mass,
the proton parton distribution functions (PDFs), and the strong coupling $\alpha_\mathrm{s}$. Furthermore, deviations from these predictions can be an indication of physics beyond the SM\@. For example, in supersymmetric (SUSY) models,
\ttbar pairs may appear as decay products
of heavier new particles, increasing the \ttbar yields.

Studies of the \ttbar production cross section, as well as dedicated searches for deviations
from the SM predictions, have been performed in recent years
by the ATLAS and CMS collaborations using a variety of production
and decay channels ~\cite{Chatrchyan:2013faa,CMStt1,Khachatryan:2015fwh,CMStopPublication2,CMStt3,CMStt4,CMStt5,top12001,CMStt7,CMStt8,CMStt9,ATLAStt1,ATLAStt2,Aad:2014kva,ATLAStopPublication5,ATLAStt4,ATLAStt5,ATLAStt7,ATLAStt8,ATLAStt9,ATLAStopPublication2, ATLAStopPublication1}. So far, all results are consistent with the SM.

This paper presents a new measurement of $\sigma_{\ttbar}$ in pp collisions at centre-of-mass energies of 7 and $8\TeV$. The measurement is performed in the \emu
channel, where each $\PW$ boson from the top quark
decays into a charged lepton and a neutrino. Compared to the previous CMS analyses in the dilepton channel at 7\TeV~\cite{top12001} and 8\TeV~\cite{Chatrchyan:2013faa}, the new measurement is performed using the complete CMS data samples recorded in the years 2011 and 2012, with integrated luminosities of 5.0 and 19.7\fbinv at $\sqrt{s} =7$ and 8\TeV, respectively. The restriction to the \emu channel
provides a pure \ttbar event sample owing to the negligible contamination from Z/$\gamma^{*}$ processes with same-flavoured leptons in the final state. The event selection is based on the kinematic properties of the leptons. An improved cross section extraction method
is used, performing
a template fit of the signal and background contributions to multi-differential
binned distributions related to the multiplicity of
b quark jets (referred to as b jets in the following) and the
multiplicity and transverse momenta of other jets in the event. The results obtained with this method (referred to as the ``reference method" in the following)
are cross-checked with an analysis performed using an event counting method.

The cross section is first determined in a fiducial (``visible'') range, $\sigma_{\ttbar}^{\text{vis}}$, defined by requirements on the transverse momentum and pseudorapidity of the electron and muon. The results are then extrapolated to obtain the cross section in the full phase space, \stt, with an additional assessment of the extrapolation uncertainties. The ratio of the cross sections at the two centre-of-mass energies is also presented. The measurements of $\stt$ at 7 and 8\TeV are used to determine, together with the NNLO prediction~\cite{mitov}, the top quark pole mass. Following a previous CMS analysis~\cite{top_alphas},
the mass is determined via the dependence of the
theoretically predicted cross section on the top quark mass.

The data are also used to constrain the cross section of pair production of the lightest supersymmetric partner of the top quark, the top squark, in the context of SUSY models with $R$-parity conservation~\cite{FARRAR1978575}. The study focuses on models predicting the decay of top squarks into a top quark and a neutralino, $\PSQt\to\PQt\PSGczDo$, and the three-body decay, $\PSQt\to\PQb\PW\PSGczDo$, with the neutralino assumed to be the
lightest supersymmetric particle (LSP)~\cite{Nilles:1983ge}. The pair production and the subsequent decays of the top squarks can lead to a final state that
is very similar to the SM \ttbar events. The search is performed with the 8\TeV data, looking for an excess of the
observed event yields of \ttbar events with respect to the SM predictions. Exclusion limits are set with 95\% confidence level (CL) for the SUSY signal strength as a function of the top squark mass for two neutralino mass hypotheses. Previous measurements setting exclusion limits in a similar regime can be found in~\cite{Aad:2014kva,Aad:2014mfk}.

This paper is structured as follows. Section~\ref{sec:cmsdet} contains a brief description of the CMS detector, followed by details of the event simulation and theoretical calculations for the \ttbar cross section are given in Section~\ref{sec:datasim}. The event selection and the definitions of the visible and total cross sections are given in Sections~\ref{sec:eventsel} and~\ref{sec:xsecdef}, respectively. The methods used to measure the cross section are explained in Section~\ref{sec:xsecext} and the systematic uncertainties are described in Section~\ref{sec:syst}. The measured \ttbar production cross sections are reported in Section~\ref{sec:results}, with the extraction of the top quark mass presented in Section~\ref{sec:mtpole}. The search for SUSY is described in Section~\ref{sec:stopsearch} and a summary is provided in Section~\ref{sec:conclusions}.

\section{The CMS detector}
\label{sec:cmsdet}
The central feature of the CMS apparatus is a superconducting solenoid of 6\unit{m} internal diameter, providing a magnetic field of 3.8\unit{T}. Within the solenoid volume are a silicon pixel and strip tracker, a lead tungstate crystal electromagnetic calorimeter (ECAL), and a brass and scintillator hadron calorimeter (HCAL), each composed of a barrel and two endcap sections. Extensive forward calorimetry complements the coverage provided by the barrel and endcap detectors. Muons are measured in gas-ionisation detectors embedded in the steel flux-return yoke outside the solenoid. A more detailed description of the CMS detector, together with a definition of the coordinate system used and the relevant kinematic variables, can be found in Ref.~\cite{JINST}.

The particle-flow (PF)~\cite{CMS-PAS-PFT-09-001, CMS-PAS-PFT-10-001} event algorithm reconstructs and identifies each individual particle with an optimised combination of information from the various elements of the CMS detector. The energy of photons is directly obtained from the ECAL measurement. The energy of electrons is determined from a combination of the electron momentum at the primary interaction vertex as determined by the tracker, the energy of the corresponding ECAL cluster, and the energy sum of all bremsstrahlung photons spatially compatible with originating from the electron track. The energy of muons is obtained from the curvature of the corresponding track. The energy of charged hadrons is determined from a combination of their momentum measured in the tracker and the matching ECAL and HCAL energy deposits, corrected for zero-suppression effects and for the response function of the calorimeters to hadronic showers. Finally, the energy of neutral hadrons is obtained from the corresponding corrected ECAL and HCAL energy.

\section{Event simulation and theoretical calculations}
\label{sec:datasim}

Experimental effects, related to the event reconstruction and choice of selection criteria, together with the detector resolution, are modelled
using Monte Carlo (MC) event generators interfaced with a detailed detector simulation. Unless specified, the same generators and parton shower models are used for the samples at 7 and 8\TeV.

The \ttbar sample is simulated using the \MADGRAPH event generator (v.~5.1.5.11)~\cite{madgraph}, which implements the relevant matrix elements at tree level with up to three additional partons. The \Madspin~\cite{bib:madspin} package is used to incorporate spin correlation effects. The value of the top quark mass is fixed to 172.5\GeV and the proton structure is described by the CTEQ6L1~\cite{bib:cteq} PDF set. The generated events are subsequently processed with \PYTHIA (v.~6.426)~\cite{Sjostrand:2006za} for parton showering and hadronisation, and the MLM prescription~\cite{bib:MLM} is used for matching of matrix-element jets to parton showers. Decays of $\tau$ leptons are handled with \TAUOLA~(v. 2.75)~\cite{tauola}.
An additional \ttbar signal sample, which is used to determine specific model uncertainties of the measurement, is obtained with the next-to-leading-order (NLO) generator \POWHEG (v.~1.0 r1380)~\cite{bib:powheg2} and also interfaced with \PYTHIA. In \POWHEG, the value of the top quark mass is also set to 172.5\GeV, and the CT10~\cite{bib:CT10} PDF set is used to describe the proton structure. The \PYTHIA Z2* tune, derived from the Z1 tune~\cite{Field:2010bc}, is used to characterise the underlying event in the \ttbar samples at 7 and 8\TeV. The Z1 tune uses the CTEQ5L PDF set, whereas Z2* adopts CTEQ6L. The propagation of the generated particles through the CMS detector and the modelling of the detector response is performed using \GEANTfour (v.~9.4)~\cite{geant}.

Only \ttbar pair decays into ${\Pe^\pm} \mu^\mp + X$ in the final state are considered signal, including intermediate leptonic $\tau$ decays. The remaining \ttbar decay modes are considered background processes and referred to as ``\ttbg.''.

The other SM background samples are simulated with \MADGRAPH (without the \Madspin package), \POWHEG, or \PYTHIA, depending on the process. The main background contributions originate from the production of W and Z/$\gamma^{*}$ bosons with additional jets (referred to in the following as W+jets and Drell--Yan (DY), respectively), single top quark $\PQt \PW$ channel, diboson ($\PW \PW$, $\PW \PZ$, and $\PZ \PZ$, referred to as ${\rm \PV \PV}$ in the following), \ttbar production in association with a Z, W, or $\gamma$ boson (referred to as ${\rm \ttbarV}$ in the following), and QCD multijet events. The W+jets, DY, and ${\rm \ttbarV}$ samples are simulated with \MADGRAPH with up to two additional partons in the final state. The \POWHEG~\cite{bib:powheg1,bib:powheg3} generator is used for simulating single top quark production, while \PYTHIA is used to simulate diboson and QCD multijet events. Parton showering and hadronisation are also simulated with \PYTHIA in all the background samples. The \PYTHIA Z2* tune is used to characterise the underlying event in the background samples at $\sqrt{s} = 8\TeV$, while the Z2 tune~\cite{bib:Z2tune} is used at $\sqrt{s} = 7\TeV$.

The simulated samples are normalised according to their expected total cross sections for integrated luminosities of 5.0\,(19.7)\fbinv for $\sqrt{s} =$~7~(8)\TeV. The expected cross sections are obtained from NNLO calculations for W+jets~\cite{Melnikov:2006di} and DY~\cite{Melnikov:2006kv} processes, NLO+next-to-next-to-leading-log (NNLL) calculations for top quark $\PQt \PW$ or ${\rm \bar{t}W}$ channel~\cite{bib:twchan}, NLO calculations for ${\rm \PV\PV}$~\cite{bib:mcfm:diboson}, $\ttbar$+$\PW$~\cite{bib:ttW}, and $\ttbar$+$\PZ$~\cite{bib:ttZ} processes, and leading-order (LO) calculations for QCD multijet events~\cite{Sjostrand:2006za}.

A number of additional pp simulated hadronic interactions (pileup) are added to each simulated event to reproduce the multiple interactions in each bunch crossing
in the data taking. The pileup events are generated using \PYTHIA. Scale factors (SFs) described in Section~\ref{sec:eventsel} are applied when needed to improve the description of the data by the simulation.

Calculations of the $\sigma_{\ttbar}$ at full NNLO accuracy in perturbative QCD, including the resummation of NNLL soft-gluon terms~\cite{topplusplus}, are used to normalise the \ttbar simulated samples and to extract the top quark pole mass. Assuming a top quark mass of $172.5\GeV$, the predicted cross sections are:
\begin{equation*}\begin{aligned}
\sigma_{\ttbar} & =   177.3^{+\,4.7}_{-\,6.0}\,\text{(scale)}\pm \phantom{0}9.0\,\text{(PDF}{+}\alpha_\mathrm{s})\unit{pb},
\quad \text{at $\sqrt{s}=7\TeV$ and} \\
\sigma_{\ttbar} & =   252.9^{+\,6.4}_{-\,8.6}\,\text{(scale)}\pm 11.7\,\text{(PDF}{+}\alpha_\mathrm{s})\unit{pb},
\quad \text{at  $\sqrt{s}=8\TeV$.}
\end{aligned}\end{equation*}

The first uncertainty is an estimate of the effect of missing higher-order corrections and is determined by independent variations of the
factorisation and renormalisation scales, $\mu_\mathrm{F}$ and $\mu_\mathrm{R}$, by factors of two, up and down from their default values (the top quark mass). The second uncertainty is associated with variations in $\alpha_\mathrm{s}$ and the PDF, following the PDF4LHC prescription with the MSTW2008 68\% CL NNLO, CT10 NNLO, and NNPDF2.3 5f FFN PDF sets (as detailed in Refs.~\cite{{pdf4lhcInterim},{pdf4lhcReport}} and references therein, as well as in Refs.~\cite{{mstw08},{Gao:2013xoa},{Ball:2012cx}}). These values were calculated using the \textsc{Top++2.0} program~\cite{topplusplus}.
The ratio of the cross sections at 7 and 8\TeV computed with NNPDF2.3,
$R_{\ttbar}^{\mathrm{NNLO}} = \sigma_{\ttbar}(8\TeV)/\sigma_{\ttbar}(7\TeV)$,
is $1.437\pm0.001\,\text{(scale)}\pm0.006\,\text{(PDF)}\pm0.001\,\text{(${\alpha}_\mathrm{s}$)}$~\cite{Czakon:2013tha}.
\section{Event selection}
\label{sec:eventsel}

At trigger level, events are required to have one electron and one muon.
For the 8\TeV data set one of the two leptons is required to have $\pt > 17\GeV$ and the other $\pt > 8\GeV$.
For the 7\TeV data set both leptons are required to have $\pt > 10\GeV$ or to fulfil the same criterion as for the 8\TeV data set.
The \emu trigger efficiency is measured in data with a method based on triggers that are uncorrelated with those used in the analysis~\cite{Chatrchyan:2013faa,bib:TOP-12-028}. In particular, the triggers require jets or missing transverse energy, which is defined as the magnitude of the projection, on the plane perpendicular to the beam direction, of the vector sum of the momenta of all reconstructed particles in an event. The trigger efficiency for events containing an \emu pair passing all selection criteria is approximately 96\% at 7\TeV and 93\% at 8\TeV.
Using the \emu trigger efficiency measured in data,
the corresponding efficiencies in the simulation are corrected
by $\eta$-dependent SFs,
which have an average value of 0.99 at 7\TeV and 0.97 at 8\TeV.

An interaction vertex~\cite{trkpas}
is required within 24\unit{cm} of the detector centre along the beam line direction,
and within 2\unit{cm} of the beam line in the transverse plane.
Among all such vertices, the primary vertex of an event
is identified as the one with the largest value
of the scalar sum of the $\pt^2$ of the associated tracks.

Leptons are required to have $\pt > 20 \GeV$
and $\abs{\eta} < 2.4$. The lepton-candidate tracks are required to originate from the primary vertex.

Lepton candidates are required to be isolated
from other PF candidates in the event.
For each electron~\cite{Khachatryan:2015hwa} or muon~\cite{Chatrchyan:2012xi} candidate, a cone with $\Delta R = 0.3$ or $0.4$, respectively,
is constructed around the track direction at the primary vertex.
Here $\Delta R$ is defined as
$\Delta R = \sqrt{\smash[b]{(\Delta \eta)^2 + (\Delta \phi)^2}}$, where
$\Delta \eta$ and $\Delta \phi$ are the differences
in pseudorapidity and azimuthal angle (in radians)
between any PF candidate and the lepton track direction.
The scalar
sum of the $\pt$ of all PF candidates
contained within the cone is calculated,
excluding the contribution from the lepton candidate itself.
All charged PF candidates not associated with the chosen primary vertex are assumed to arise from pileup events, and are excluded from the calculation of the \pt deposited in the cone. The neutral component is also corrected for pileup effects.
The relative isolation discriminant, $I_\text{rel}$, is defined as the ratio of this sum to the $\pt$ of the lepton candidate.
An electron candidate is selected if $I_\text{rel}<0.10$; the corresponding requirement for muons is $I_\text{rel}<0.12$.

The efficiency of the lepton selection is measured using a ``tag-and-probe'' method
in dilepton events enriched with $\PZ$ boson candidates~\cite{inclusWZ3pb,top12001}.
The measured values for the combined identification and isolation efficiencies are
typically 80\% for electrons and 90\% for muons.
The lepton identification efficiencies in simulation are corrected to the measured values
in data by $\pt$ and $\eta$ dependent SFs, which have values in the range 0.97--0.99. From all events that contain oppositely charged lepton pairs, events are selected if the lepton pair with the largest value of the scalar sum of the \pt corresponds to an \emu pair. Candidate events with \emu invariant masses $m_{\Pe\Pgm} <20\GeV$ are removed to reduce the contamination from QCD multijet processes. This selection is referred to as ``\emu selection''.

Jets are reconstructed
using the  anti-\kt clustering algorithm~\cite{antikt} with a distance parameter $R=0.5$.
The algorithm uses the PF candidates as input objects.
To minimise the impact of pileup, charged particle candidates not associated with the
primary vertex are excluded.
The jet energy is corrected for pileup in a manner similar to the correction of the total energy inside the lepton
isolation cone.
Additional jet energy corrections are also applied as a function of the jet
$\pt$ and $\eta$~\cite{bib:JME-10-011:JES}.
Jets are selected if they have $\pt> 30\GeV$ and $\abs{\eta}< 2.4$ and the angular distance between them and the
selected leptons satisfies $\Delta R({\rm jet,lepton})>0.5$.

As the \ttbar events are expected to contain mainly jets from the hadronisation of \PQb quarks, requiring the presence of b jets can reduce background from events without b quarks.
Jets are identified as b jets (b-tagged)  using the combined secondary vertex algorithm~\cite{BTV-11-004-pub}. The discriminator threshold chosen for the reference method to extract the cross section corresponds
to an identification efficiency for b jets of about 50\%
and a misidentification (mistag) probability
of about 10\% for c quark jets and
0.1\% for light-flavour jets (u, d, s, and gluons).
A looser discriminator threshold is chosen for the event counting method such that the efficiency is about 70\% for jets originating from b quarks and 20\% for c quark jets,
while the probability of mistagging for jets originating
from light flavours is around 1\%~\cite{BTV-11-004-pub}. For the reference method there are no constraints on the number of jets and b-tagged jets in the event.

\begin{figure}[htbp]
\centering
\includegraphics[width=0.495\textwidth]{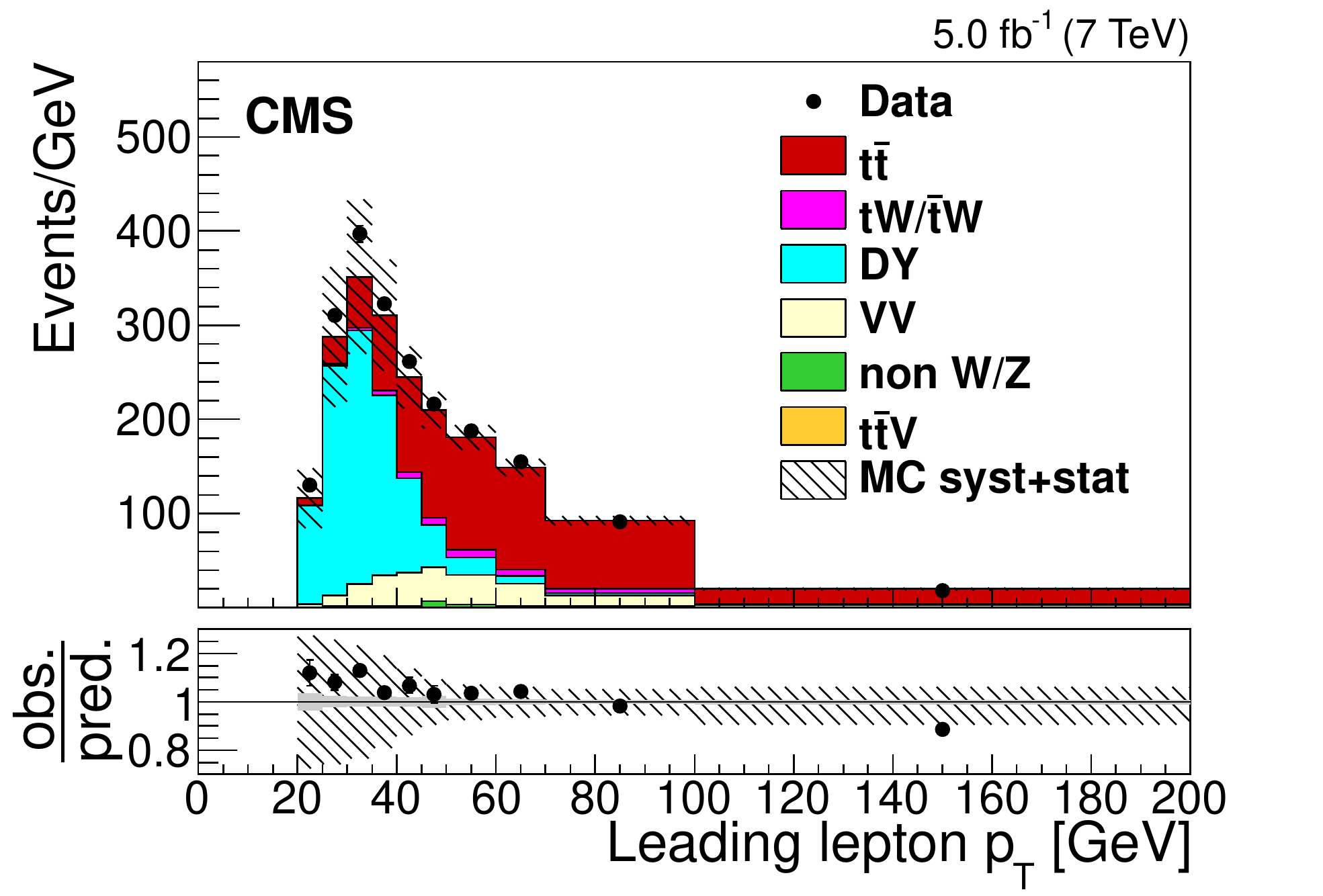}
\includegraphics[width=0.495\textwidth]{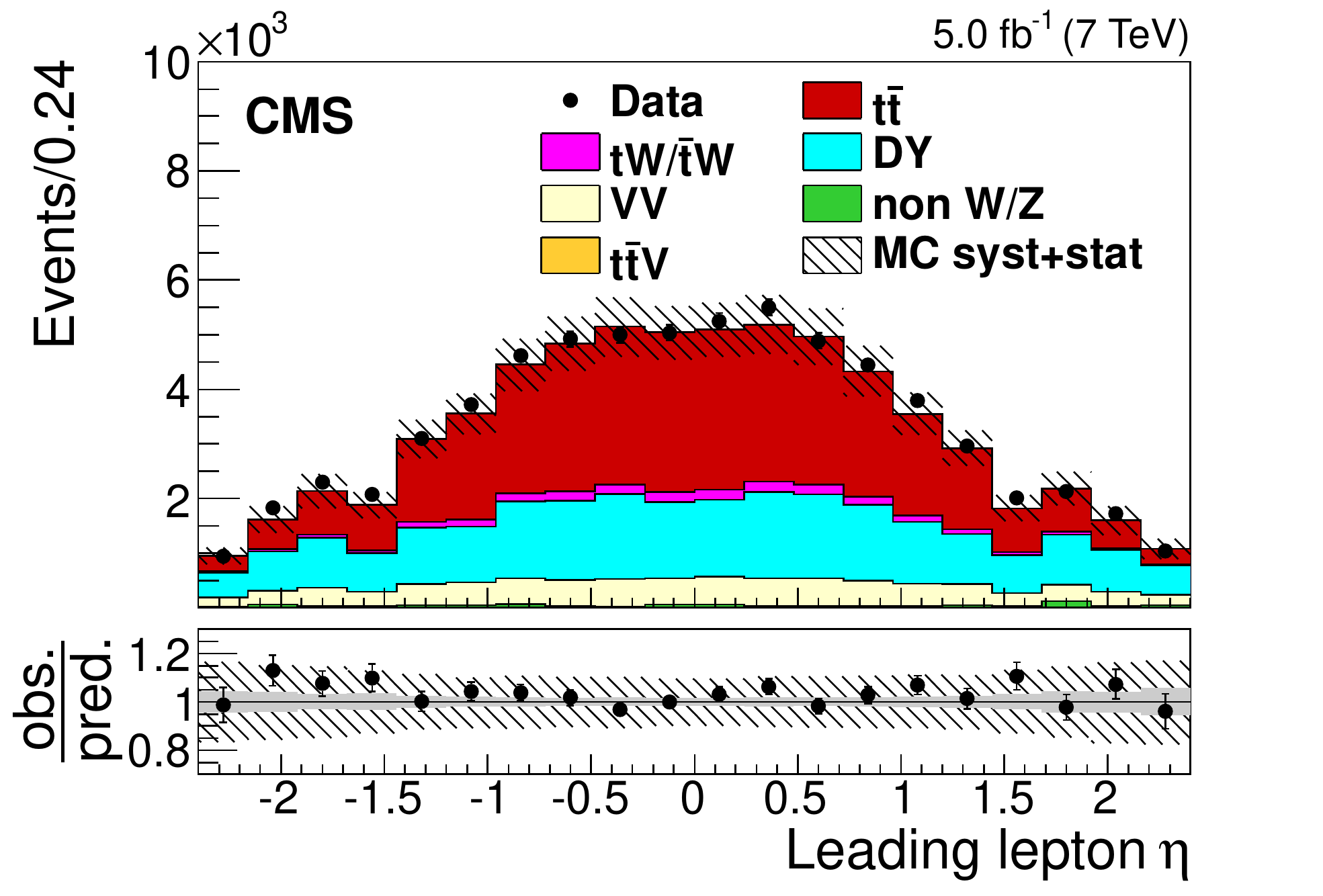}
\includegraphics[width=0.495\textwidth]{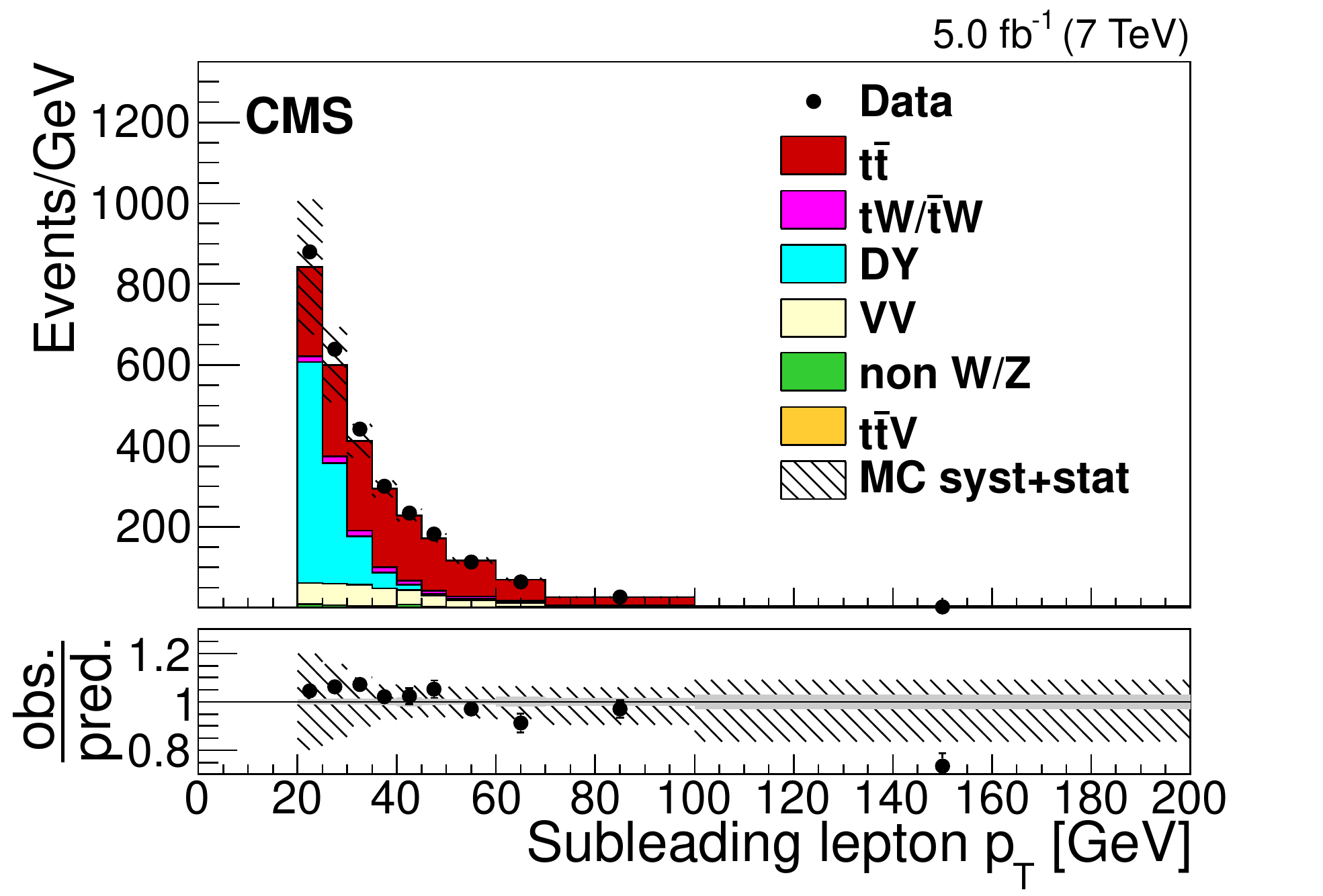}
\includegraphics[width=0.495\textwidth]{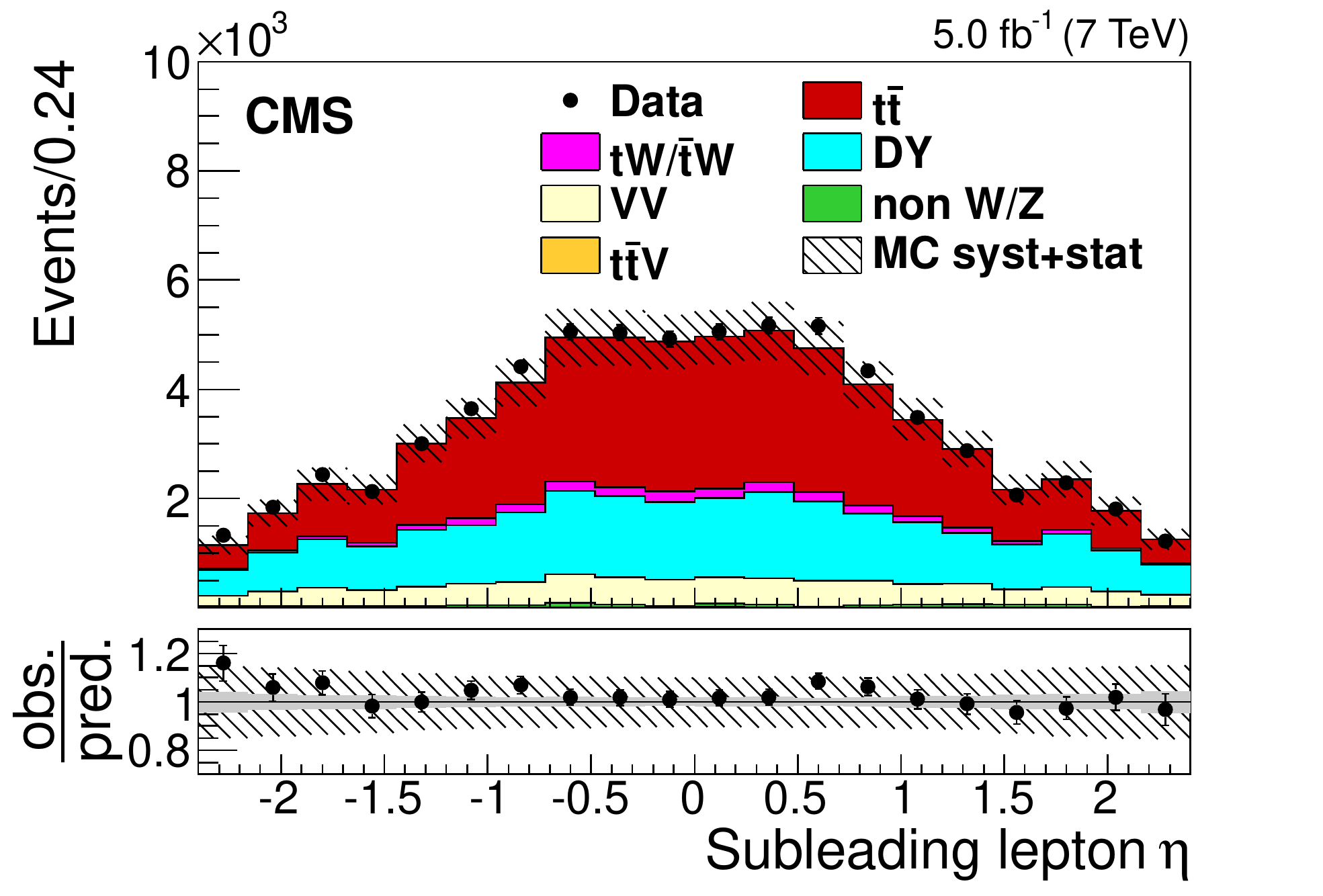}
\caption{Distributions of \pt (left) and $\eta$ (right) of the leading (top) and subleading (bottom) leptons, after the \emu selection, for the 7\TeV data. The last bin of the \pt distributions includes the overflow events. The hatched bands correspond to the total uncertainty in the sum of the predicted yields. The ratios of data to the sum of the predicted yields are shown at the bottom of each plot. Here, an additional solid gray band represents the contribution from the statistical uncertainty in the MC simulation. The contributing systematic uncertainties are discussed in Section~\ref{sec:syst}.
\label{fig:lh_ctrplots_7TeV}}
\end{figure}

\begin{figure}[htbp]
\centering
\includegraphics[width=0.495\textwidth]{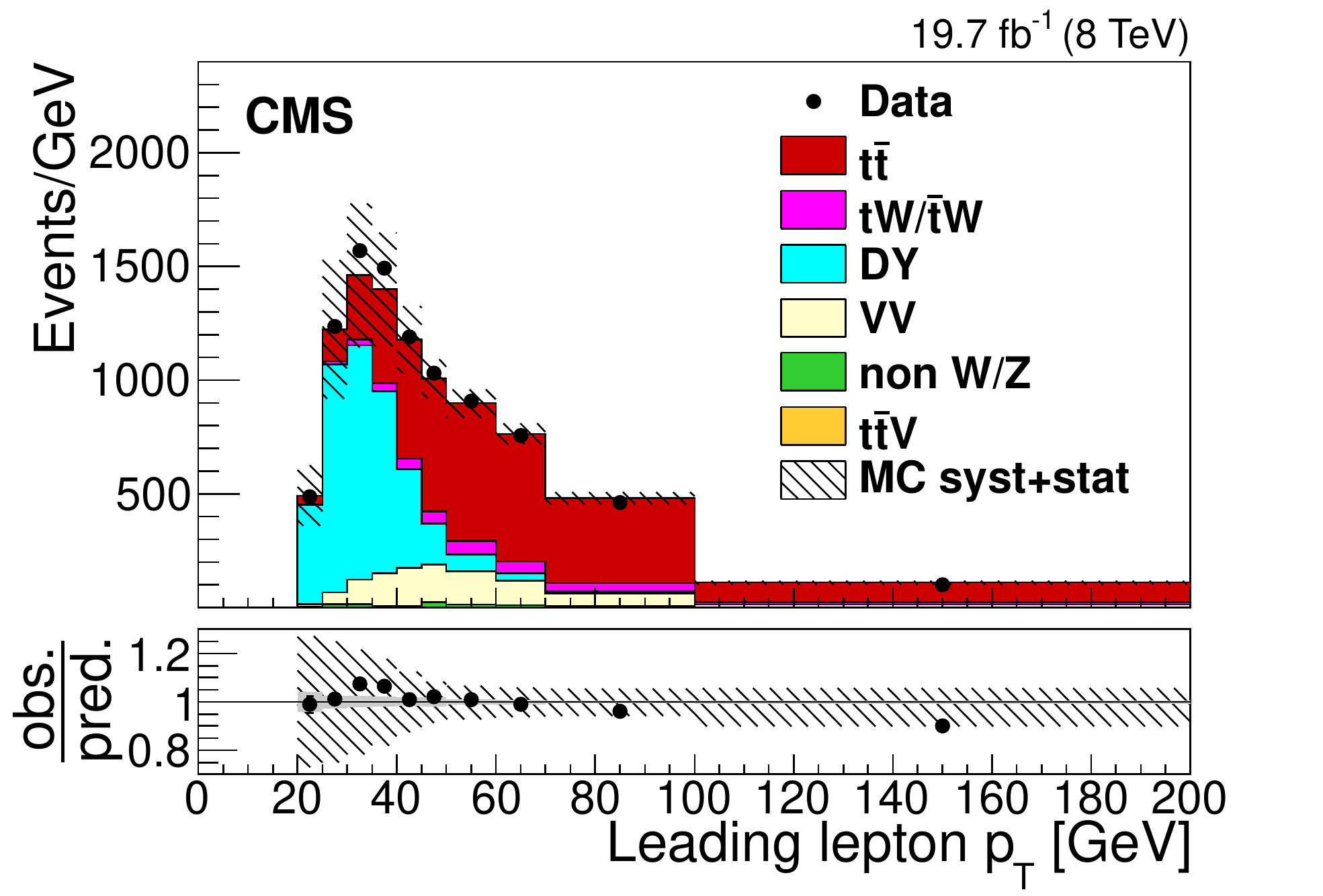}
\includegraphics[width=0.495\textwidth]{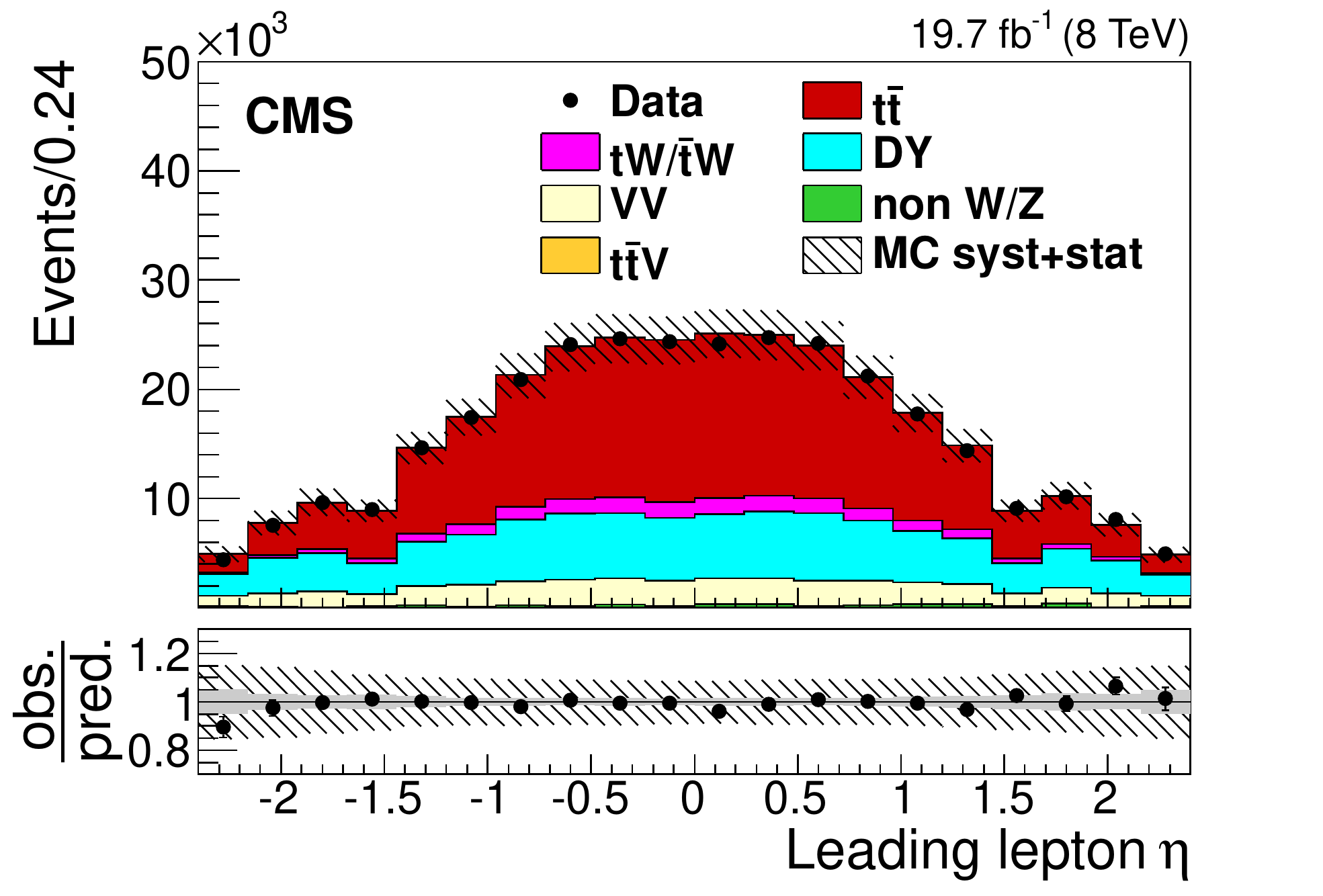}
\includegraphics[width=0.495\textwidth]{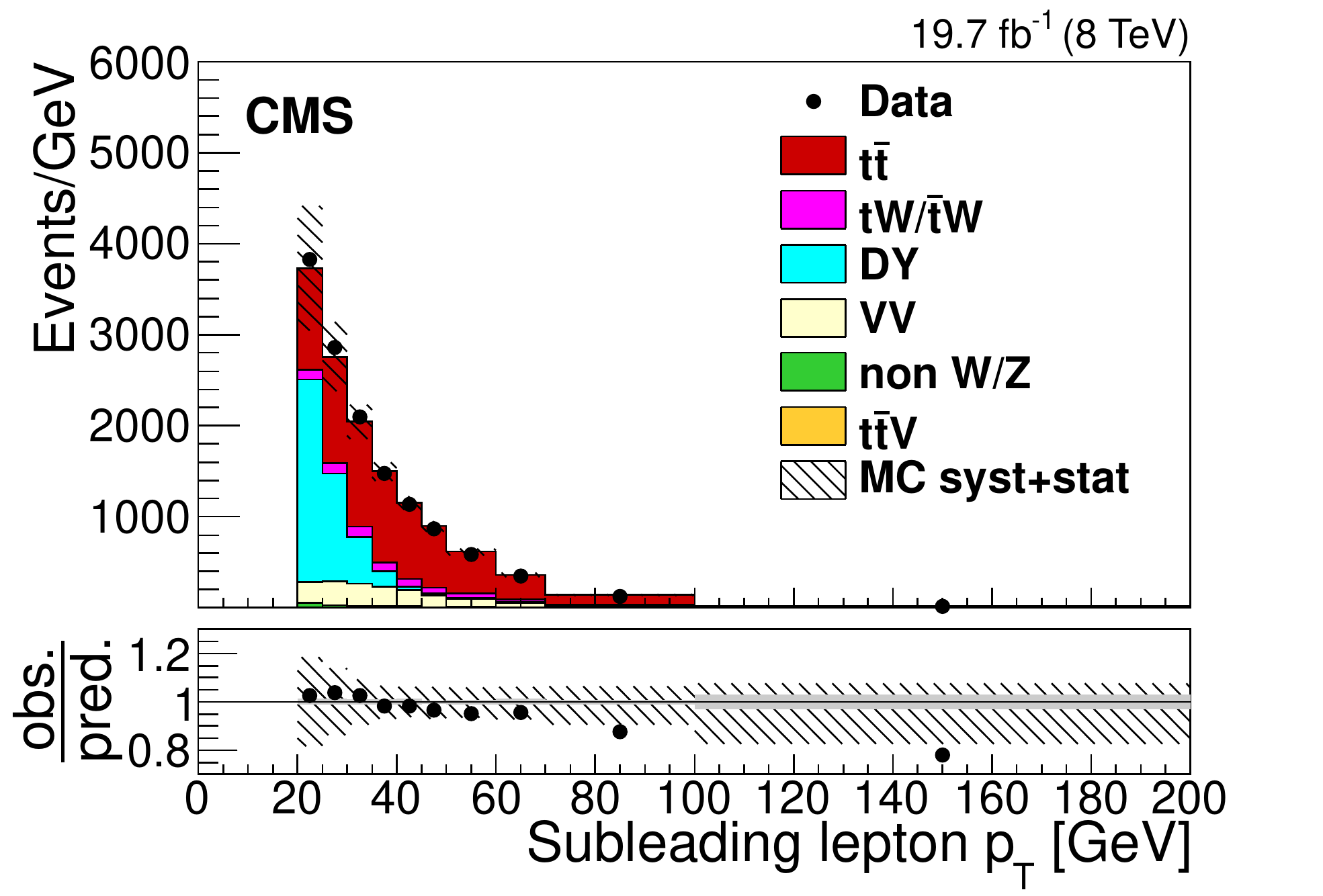}
\includegraphics[width=0.495\textwidth]{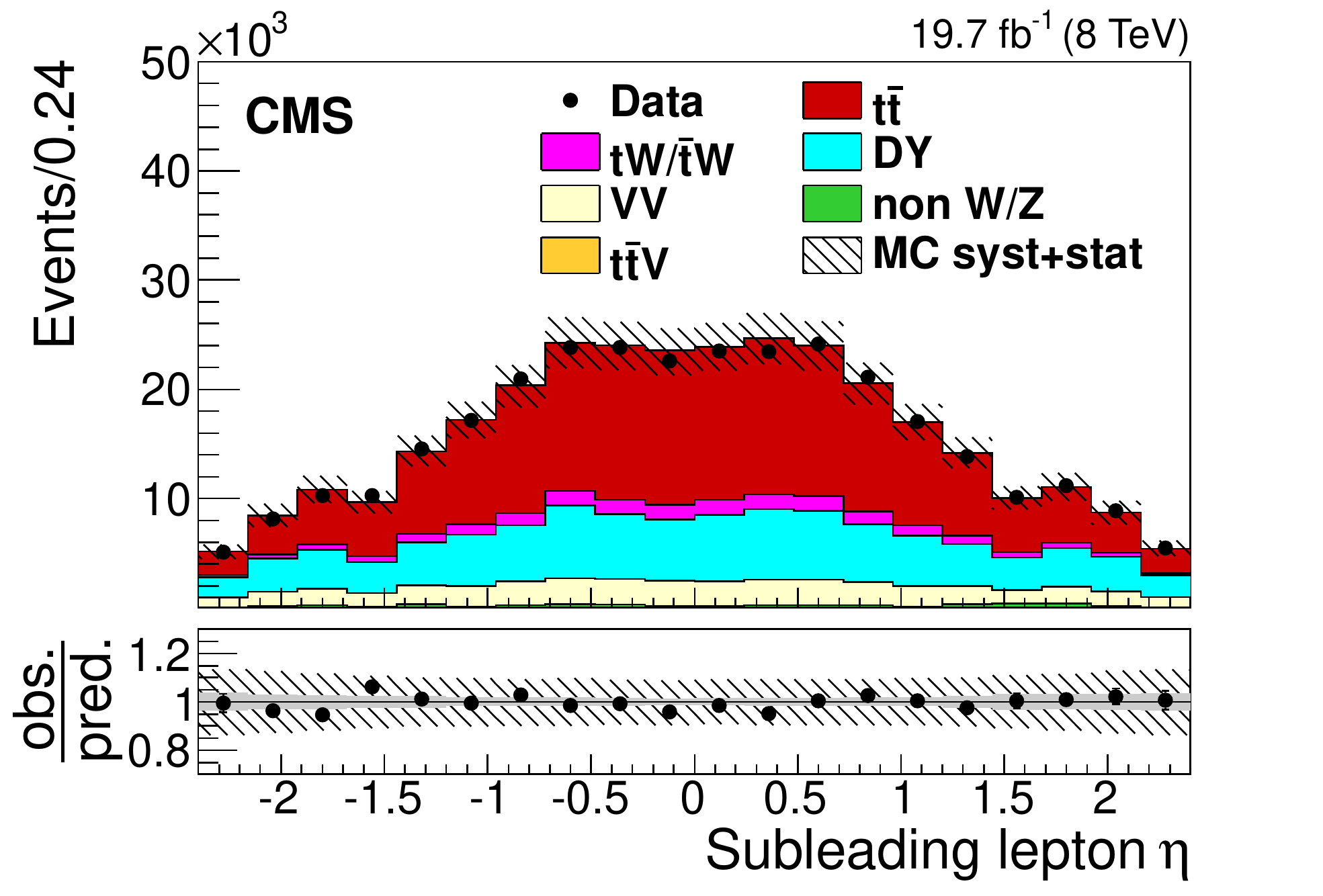}
\caption{Distributions of \pt (left) and $\eta$ (right) of the leading (top) and subleading (bottom) leptons, after the \emu selection, for the 8\TeV data. The last bin of the \pt distributions includes the overflow events. The hatched bands correspond to the total uncertainty in the sum of the predicted yields. The ratios of data to the sum of the predicted yields are shown at the bottom of each plot. Here, an additional solid grey band represents the contribution from the statistical uncertainty in the MC simulation. The contributing systematic uncertainties are discussed in Section~\ref{sec:syst}.
\label{fig:lh_ctrplots_8TeV}}s
\end{figure}

Figures~\ref{fig:lh_ctrplots_7TeV} and~\ref{fig:lh_ctrplots_8TeV} show for the 7 and 8\TeV data and simulations, respectively, the $\pt$ and $\eta$ distributions of the highest (leading)
and second-highest (subleading) $\pt$ lepton from the selected
\emu pair, after the \emu selection is applied. The data are compared to the expected distributions for the \ttbar signal and individual backgrounds, which are derived from MC simulated samples.
The contributions from QCD multijet, W+jets, and \ttbar background processes arise from events where at least one jet is incorrectly reconstructed as a lepton or a lepton that does not originate from a prompt $\PW$ or $\PZ$ boson decay fulfils the selection criteria. These contributions are referred to as ``non W/Z'' background.

In general, the sum of the estimated contributions provides an adequate description of the data, within uncertainties. However, as observed previously~\cite{bib:TOP-12-028}, the simulation is seen to have a somewhat harder \pt spectrum than measured. The impact on the measurement is accounted for by including an additional modelling uncertainty.


\begin{figure}[htbp]
\centering
\includegraphics[width=0.495\textwidth]{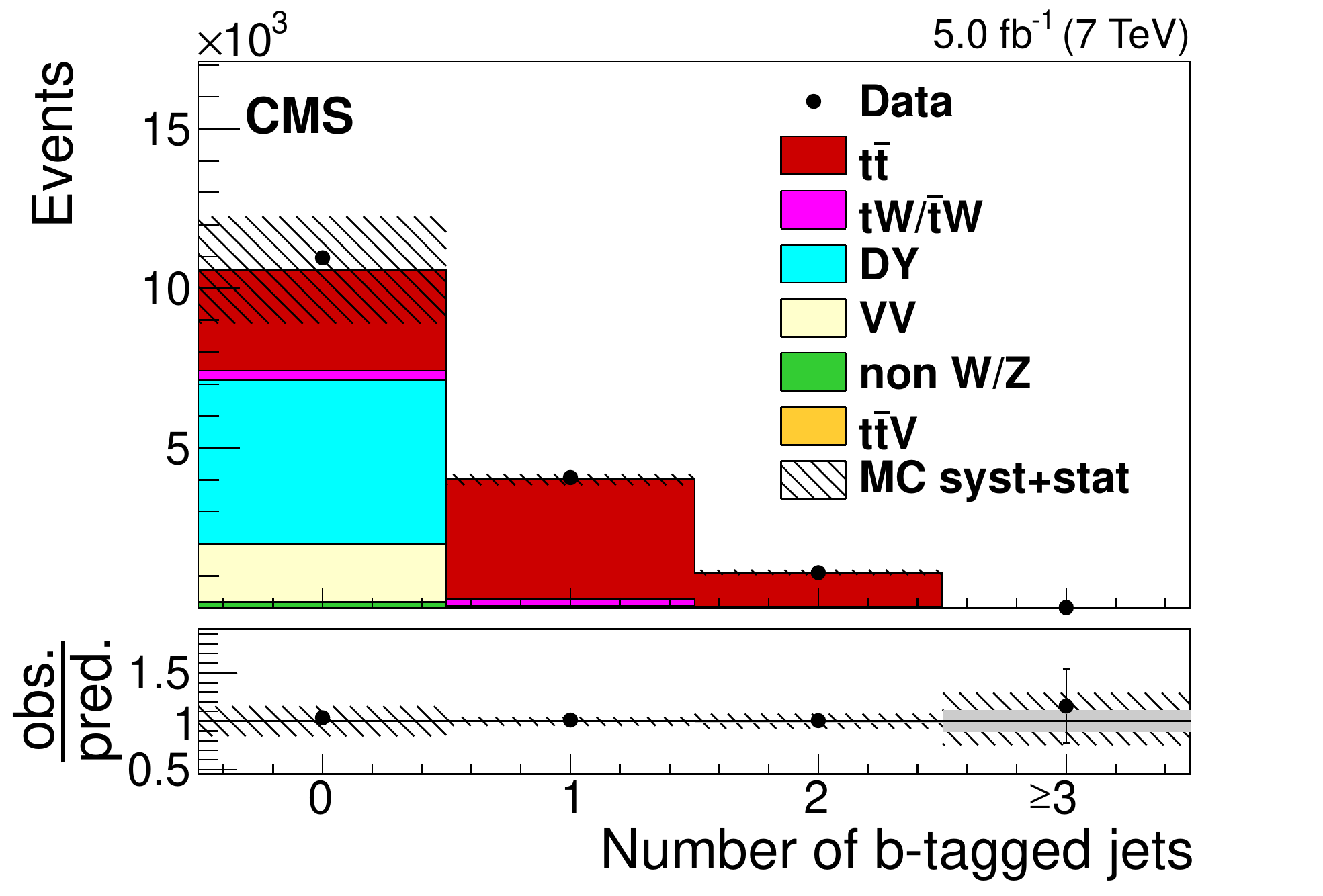}
\includegraphics[width=0.495\textwidth]{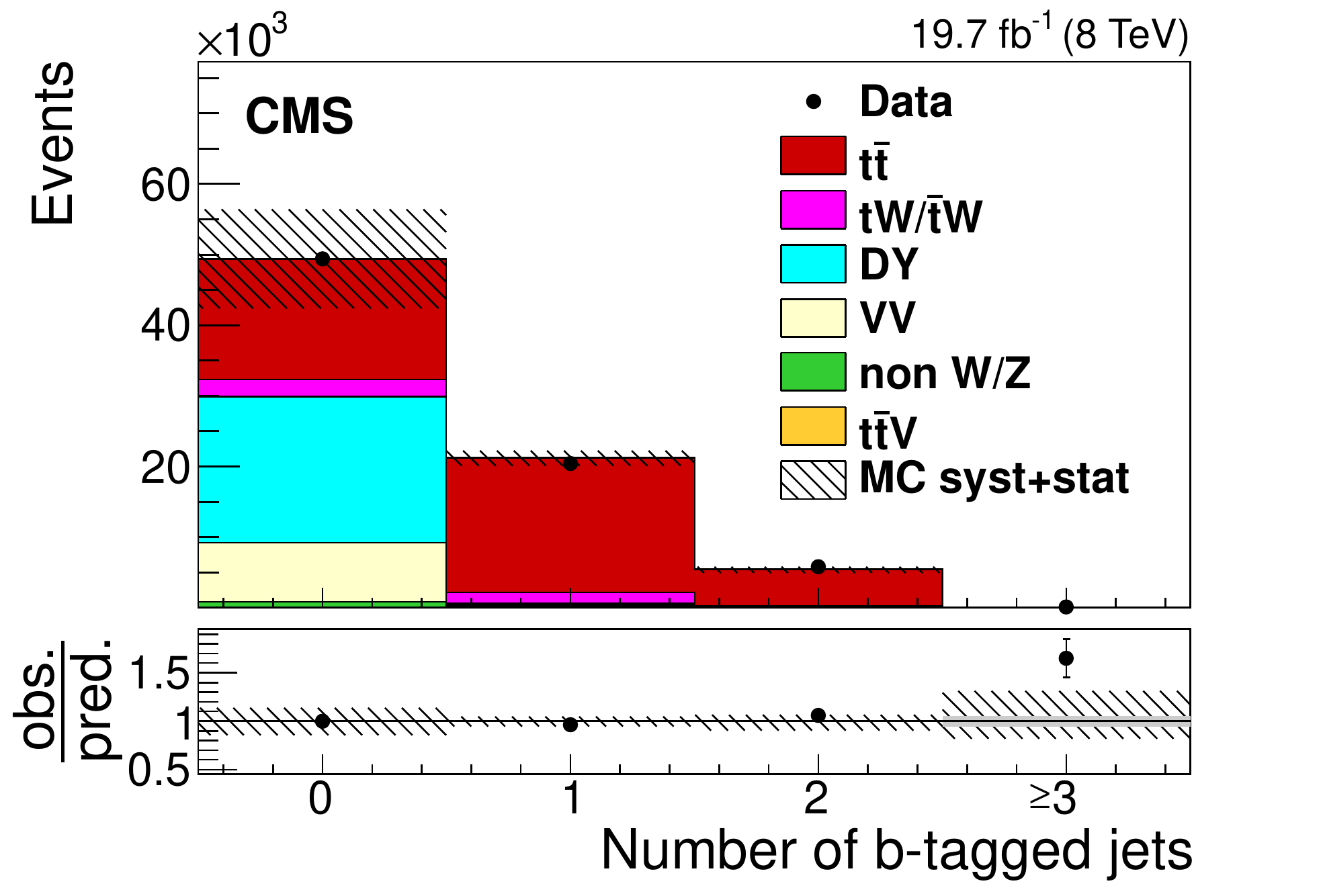}
\caption{Number of b-tagged jets after the \emu selection for 7\TeV (left) and 8\TeV (right). The hatched bands correspond to the total uncertainty in the sum of the predicted yields. The ratios of data to the sum of the predicted yields are shown at the bottom of each plot. Here, an additional solid grey band represents the contribution from the statistical uncertainty in the MC simulation. The contributing systematic uncertainties are discussed in Section~\ref{sec:syst}.
}
       \label{fig:lh_btagmulti}
\end{figure}

Figure~\ref{fig:lh_btagmulti} shows the number of b-tagged jets in events passing the \emu selection at 7 and 8\TeV. It should be noted that the size of the uncertainties in Figs.~\ref{fig:lh_ctrplots_7TeV}--\ref{fig:lh_btagmulti} does not reflect those in the final measurements, which are constrained by the likelihood fit described in Section~\ref{sec:xsect_binlh}.
Good agreement is observed between data and the sum of the expected yields.

\section{Cross section definitions}
\label{sec:xsecdef}
The \ttbar production cross sections are first measured in a fiducial range,
defined within the kinematic acceptance of the \ttbar decay particles
that are reconstructable in the detector. This avoids the need for extrapolating the cross sections
into the unmeasured kinematic phase space of these particles. In this analysis the fiducial range is
defined by the $\pt$ and $\eta$ requirements on the electron and muon
in the final state. The visible cross section, $\sigma^{\text{vis}}_{\ttbar}$, is defined for events containing an
oppositely charged \emu pair from the decay chain
${\rm t} \to {\rm W b} \to {\ell} \nu {\PQb}$ (including $\PW \to \tau \nu \to {\ell} \nu \nu \nu$)
and with both leptons satisfying $\pt > 20\GeV$ and $\abs{\eta} < 2.4$.
This visible cross section is then extrapolated to
obtain the cross section for \ttbar production at parton level
in the full phase space using the formula
\begin{equation}
\stt =  \frac{\sigma^{\text{vis}}_{\ttbar}}{A_{\Pe\mu}}.
\label{eq:extrapol}
\end{equation}
Here, $A_{\Pe\mu}$ denotes the acceptance defined as the
fraction of all \ttbar events fulfilling the
above selection criteria for the visible cross section.
The acceptance is determined from the simulated
\ttbar signal sample, and includes the leptonic branching fraction of the $\PW$ bosons of 10.86\%~\cite{Agashe:2014kda}.

\section{Analysis methods for the measurement of the cross section}
\label{sec:xsecext}
Two methods are used to measure the \ttbar production cross section. The reference method is a binned likelihood fit to multi-differential final state distributions, performed in categories of number of additional and b-tagged jets, as described in Section~\ref{sec:xsect_binlh}. In addition, an analysis is performed using an event counting technique, as explained in Section~\ref{sec:xsecext_evc}.

\subsection{Binned likelihood fit}
\label{sec:xsect_binlh}
An extended binned likelihood fit is applied to
determine $\sigma_{\ttbar}^{\text{vis}}$. The expected signal and background distributions
are modelled in the fit by template histograms constructed
from the simulated samples. The free parameters in the fit
are $\sigma^{\text{vis}}_{\ttbar}$, the background normalisation
parameters $\vec{\omega} = (\omega_1, \omega_2, \ldots, \omega_{K})$
for the $K$ sources of backgrounds, and the $M$
nuisance parameters $\vec{\lambda} = (\lambda_1, \lambda_2, \ldots, \lambda_M)$,
representing sources of systematic uncertainties other than the background normalisation, such as the
jet energy scale and the trigger efficiency.
The likelihood function $L$, based on Poisson statistics,
is given by
\begin{equation}
L  =  \prod_{i} \left( \exp{\left[- \mu_i \right]}
\mu_i^{n_i} /n_i! \right) \, \prod_{k = 1}^{K} \pi(\omega_k)\, \prod_{m = 1}^{M} \pi(\lambda_m).
\label{eq:lhfunct}
\end{equation}
Here, $i$ denotes the bin index of the chosen final state distribution,
and $\mu_i$ and $n_i$ are the expected and observed event numbers
in bin $i$.
The terms $\pi(\omega_k)$ and $\pi(\lambda_m)$
denote prior probability density functions for the
background and the other nuisance parameters, representing
the prior knowledge of these parameters.
The Poisson expectation values $\mu_i$ can be further decomposed as
\begin{equation}
\mu_i = s_i(\sigma^{\text{vis}}_{\ttbar},\vec{\lambda})
+ \sum_{k = 1}^{K} b_{k,i}^{\mathrm{MC}}(\vec{\lambda}) \, (1 + \gamma_k \omega_k).
\label{eq:expectev}
\end{equation}
Here, $s_i$ denotes the expected number of
\ttbar signal events, which depends on $\sigma^{\text{vis}}_{\ttbar}$
and the nuisance parameters
$\vec{\lambda}$.
The quantity $b_{k,i}^{\mathrm{MC}}$ represents the nominal
template prediction of background events from source $k$ in bin $i$, and $\gamma_k$ its estimated relative global normalisation uncertainty. In this analysis the background normalisation parameters
$\omega_k$ and the other nuisance parameters $\lambda_m$
are defined such that
each prior can be represented by a unit normal distribution, unless mentioned otherwise.

A suitable differential distribution for the likelihood fit is
the number of selected b-tagged jets in the event. The probability to reconstruct and identify one of the two b jets from the decaying \ttbar pair is nearly independent of the probability to reconstruct and identify the other b jet. Because of the large mass of the top quark, the kinematic properties of the two b jets are determined to a large extent by the nearly independent decay topologies of the t and $\rm{\bar{t}}$, and strong kinematic acceptance correlations arise only for extreme production topologies, such as for \ttbar pairs with a large Lorentz boost.

Under the assumption of the independence of the probabilities to identify the b jets, it is possible to express the number of expected signal events with exactly one ($s_1$), and exactly two ($s_2$) b-tagged jets using binomial probabilities~\cite{Aad:2014kva}:
\begin{align}
s_1  & =  s_{\emu}  \, 2 \epsilon_{\PQb}(1-C_{\PQb}\epsilon_{\PQb}),  \label{eq:nb1} \\
s_2  & =  s_{\emu}  \,   \epsilon_{\PQb}^2 C_{\PQb}. \label{eq:nb2}
\end{align}
Here, $s_{\emu}$ is the total number of events after the \emu selection and can be written as $s_{\emu} = \mathcal{L} \sigma^{\text{vis}}_{\ttbar}\epsilon_{\emu}$, with $\mathcal{L}$ being the integrated luminosity and $\epsilon_{\emu}$ the efficiency for events to pass the \emu selection. The parameter $\epsilon_{\PQb}$ comprises the total efficiency that a b jet is reconstructed within the kinematic acceptance and b-tagged.
The quantity $C_{\PQb}$ corrects for the small correlations
between the tagging of the two b jets and can be expressed as
$C_{\PQb}=4s_{\emu}s_2/
(s_1+2s_2)^2$.

The remaining signal events with zero or more than two b-tagged jets are considered in a third category:
\begin{equation}
s_{0} = s_{\emu} \, \left[ 1-2\epsilon_{\PQb}(1-C_{\PQb}\epsilon_{\PQb})-C_{\PQb}\epsilon_{\PQb}^2\right].
\label{eq:nb0}
\end{equation}

In Ref.~\cite{Aad:2014kva}, two equations similar to Eqs.~(\ref{eq:nb1},~\ref{eq:nb2}) are directly solved for the \ttbar production cross section
and $\epsilon_{\PQb}$.
In the present analysis, Eqs.~(\ref{eq:nb1},~\ref{eq:nb2}) are used together with Eq.~(\ref{eq:nb0})
in the template fit. The quantities $\epsilon_{\emu}$, $\epsilon_{\PQb}$, and $C_{\PQb}$ are directly determined
from the \ttbar signal simulation, expressing
$\epsilon_{\PQb}$\/ as ${(s_1 + 2 s_2)}/{2s_{\emu}}$, and parametrised as a function of the nuisance parameters $\vec{\lambda}$. The nominal values for the 8\TeV simulated \ttbar signal are $\epsilon_{\emu}=0.51$, $\epsilon_{\PQb}=0.36$, and $C_{\PQb}=0.99$, and the values for the 7\TeV sample are similar. The use of these equations facilitates an accurate modelling
of the expected signal rates as a function of the nuisance parameters,
\ie avoiding mismodelling effects that could arise from approximating
the dependences as linear functions.

In order to improve the sensitivity of the fit,
the events are further categorised
into four classes of multiplicity of additional jets
in the event (zero, one, two, and three or more additional jets). This leads, together with the three classes of b-tagged jets,
to 12 different categories in total. Additional jets must be non-b-tagged jets.
In case there is no additional jet, the corresponding
event yields are directly used in the likelihood fit,
otherwise events are further categorised into bins of the
\pt of the least energetic additional jet in the event.

The signal subcategory probabilities, background rates, and values of $\epsilon_{\emu}$, $\epsilon_{\PQb}$, and $C_{\PQb}$ are obtained from simulation and depend on the nuisance parameters $\vec{\lambda}$.
Each relevant dependency of a quantity
on a parameter $\lambda_m$ is modelled by a second-order polynomial,
that is constructed from evaluating the quantity at three values $\lambda_m=0,1,-1$,
corresponding to the nominal value of the parameter and to ${\pm}1$ standard deviation ($\sigma$) variations.
For a few sources of uncertainty, only one exact variation is possible, \eg
when there are only two variants of signal generators available that differ
in a certain uncertainty source such as the matrix element calculation;
in such cases, a linear function is chosen to model the dependence
of the quantity on the respective $\lambda_m$.
For several nuisance parameters
representing systematic modelling uncertainties in the measurement,
a box prior is chosen instead of the standard unit normal prior,
with a value of 0.5 between ${-}1$ and ${+}1$ and zero elsewhere.
Such priors are chosen for the following uncertainties (discussed in Section~\ref{sec:syst_model}):
renormalisation and factorisation scales, jet-parton matching scale, top quark \pt modelling,
colour reconnection, underlying event, and matrix element generator.

\begin{figure}[htbp]
\centering
\includegraphics[width=\textwidth]{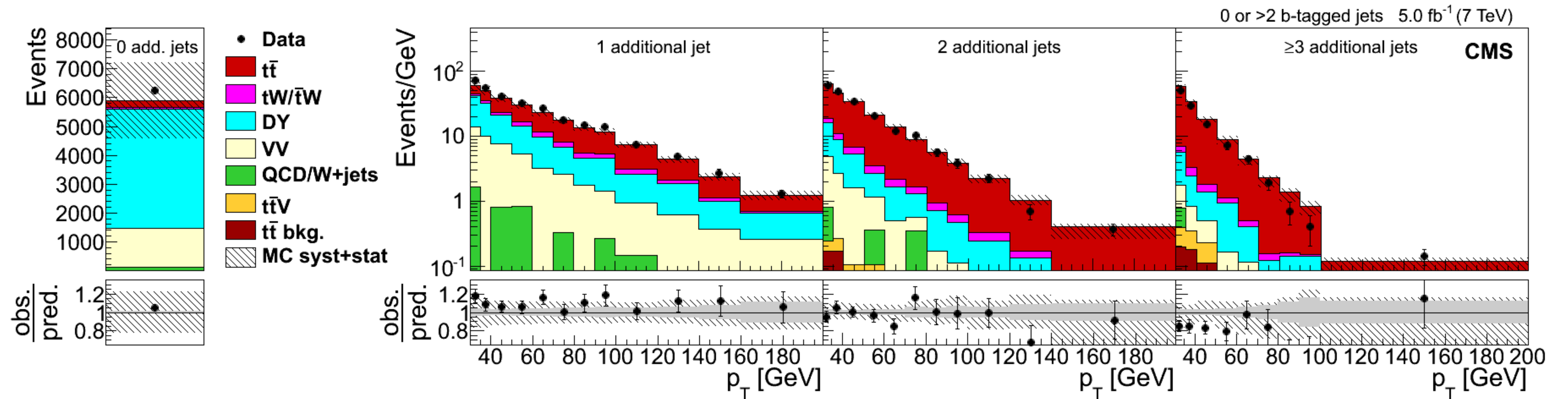}
\includegraphics[width=\textwidth]{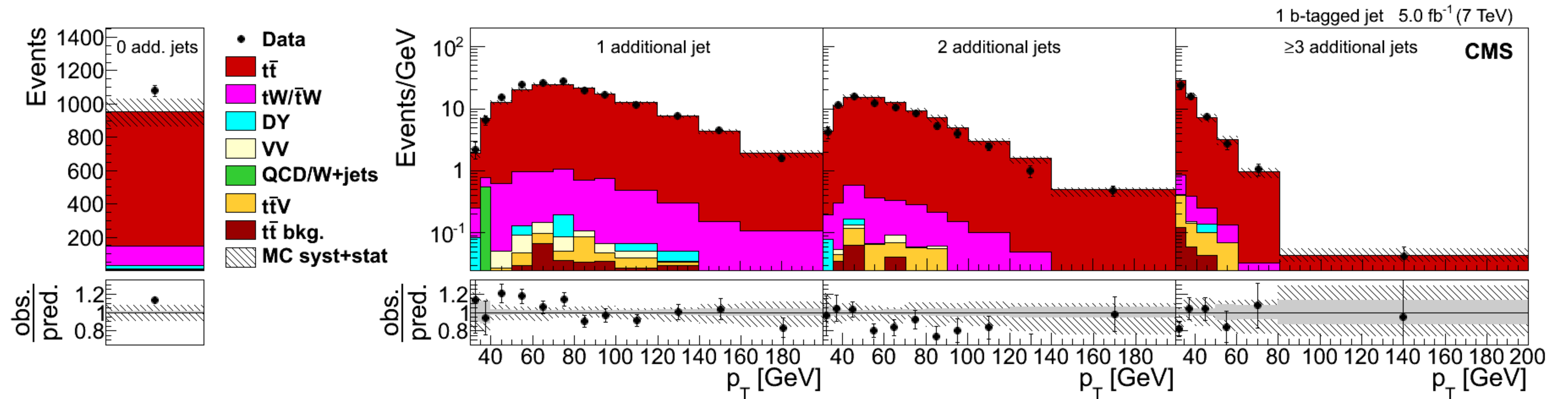}
\includegraphics[width=\textwidth]{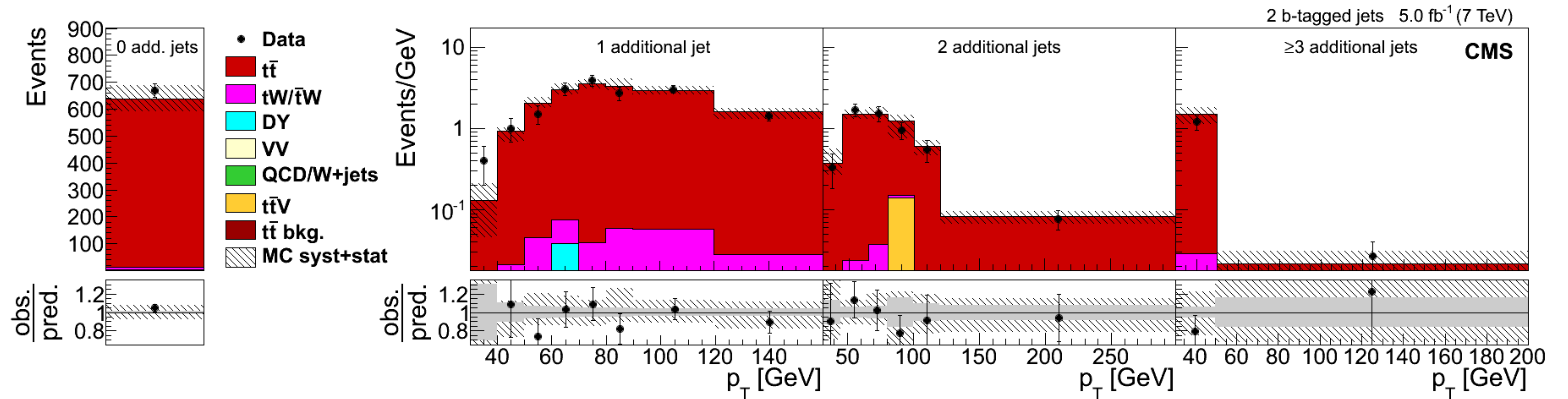}
\caption{Total event yield for zero additional non-b-tagged jets (left)
and \pt of the non-b-tagged jet with the lowest \pt in the event (right) for events with one, two, and at least three additional non-b-tagged jets, and with zero or more than two (top row), one (middle row), and two (bottom row) b-tagged jets at $\sqrt{s} = 7\TeV$. The last bin of the \pt distributions includes the overflow events. The hatched bands correspond to the sum of statistical and systematic uncertainties in the event yield for the sum of signal and background predictions. The ratios of data to the sum of the predicted yields are shown at the bottom of each plot. Here, an additional solid grey band represents the contribution from the statistical uncertainty in the MC simulation.
 \label{fig:lh_inputdistr_7TeV}}
\end{figure}

\begin{figure}[htbp]
\centering
\includegraphics[width=\textwidth]{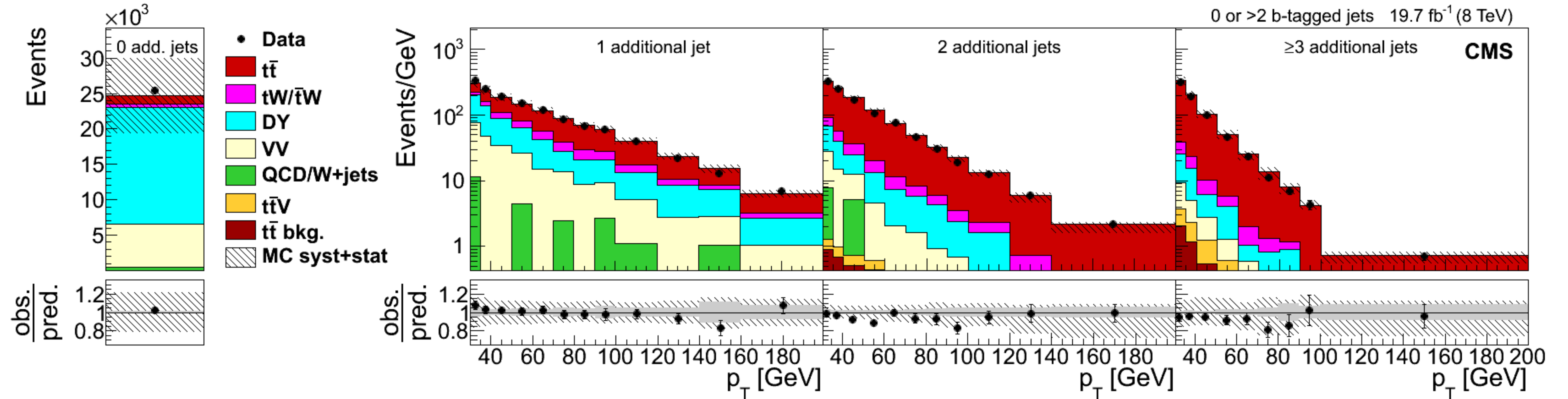}
\includegraphics[width=\textwidth]{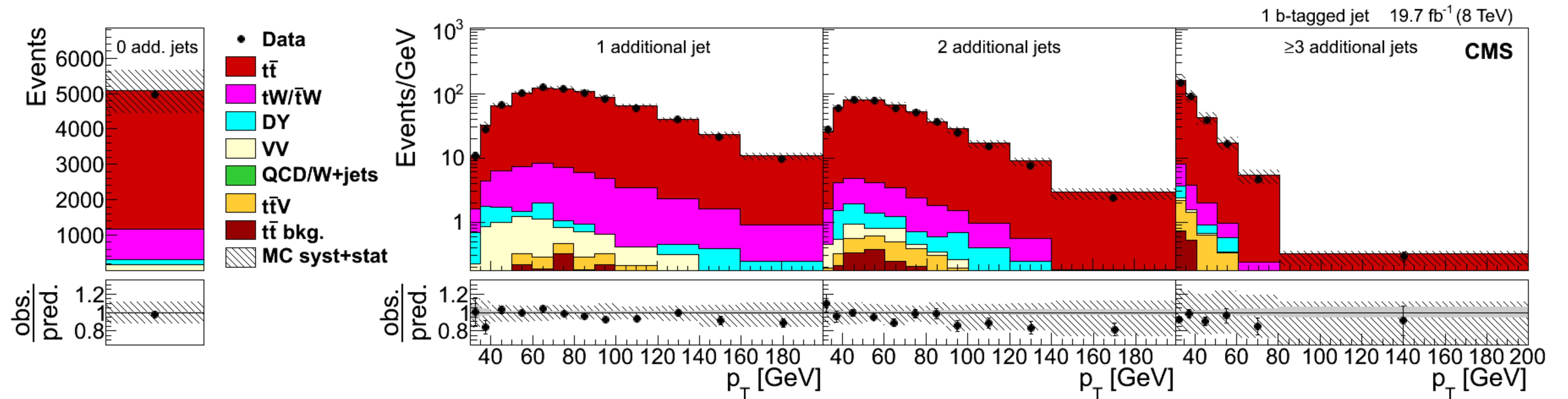}
\includegraphics[width=\textwidth]{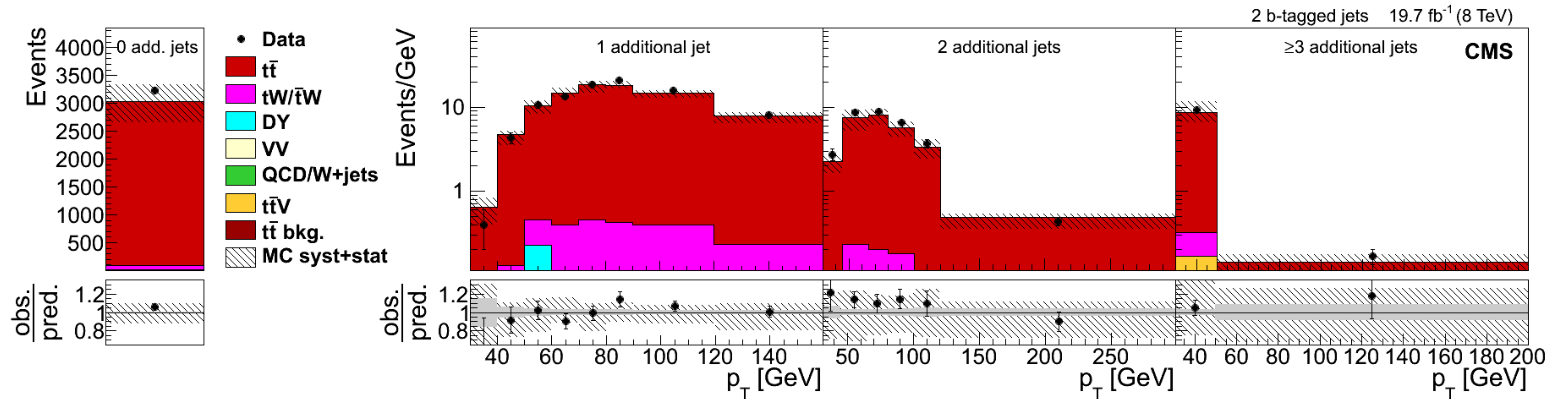}
\caption{Total event yield for zero additional non-b-tagged jets (left)
and \pt of the additional non-b-tagged jet with the lowest \pt in the event (right) for events with one, two, and at least three additional non-b-tagged jets, and with zero or more than two (top row), one (middle row), and two (bottom row) b-tagged jets at $\sqrt{s} = 8\TeV$. The last bin of the \pt distributions includes the overflow events. The hatched bands correspond to the sum of statistical and systematic uncertainties in the event yield for the sum of signal and background predictions. The ratios of data to the sum of the predicted yields are shown at the bottom of each plot. Here, an additional solid grey band represents the contribution from the statistical uncertainty in the MC simulation.
       \label{fig:lh_inputdistr_8TeV}}
\end{figure}

The likelihood fit is finally performed using the
function
$\chi^2 = -2 \ln L$, where $L$ is the likelihood function
given in Eq.~(\ref{eq:lhfunct}).
The \textsc{minuit}~\cite{James:1975dr}
program is used to minimise this $\chi^2$ as function
of the free fit parameters $\stt$, $\vec{\omega}$, and $\vec{\lambda}$.
The fit uncertainty in $\stt$ is determined
using \textsc{minos}, the profile likelihood algorithm which is part of \textsc{minuit}. Figures~\ref{fig:lh_inputdistr_7TeV} and~\ref{fig:lh_inputdistr_8TeV} show
the multi-differential distributions used in the fit.
A reasonably good agreement is found between data and expectations before the fit.


\begin{figure}[htbp]
\centering
\includegraphics[width=\textwidth]{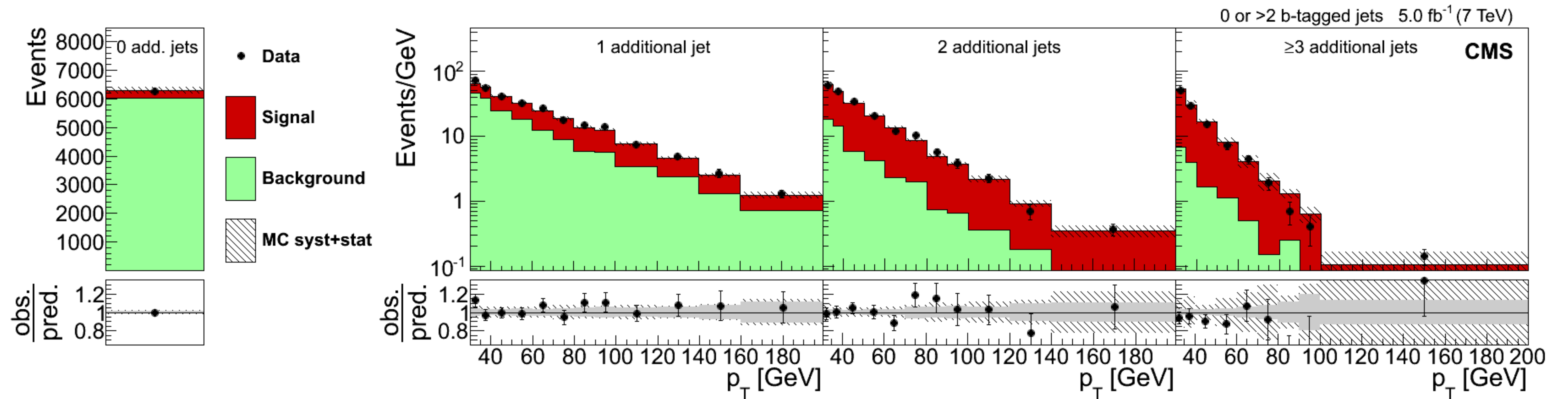}
\includegraphics[width=\textwidth]{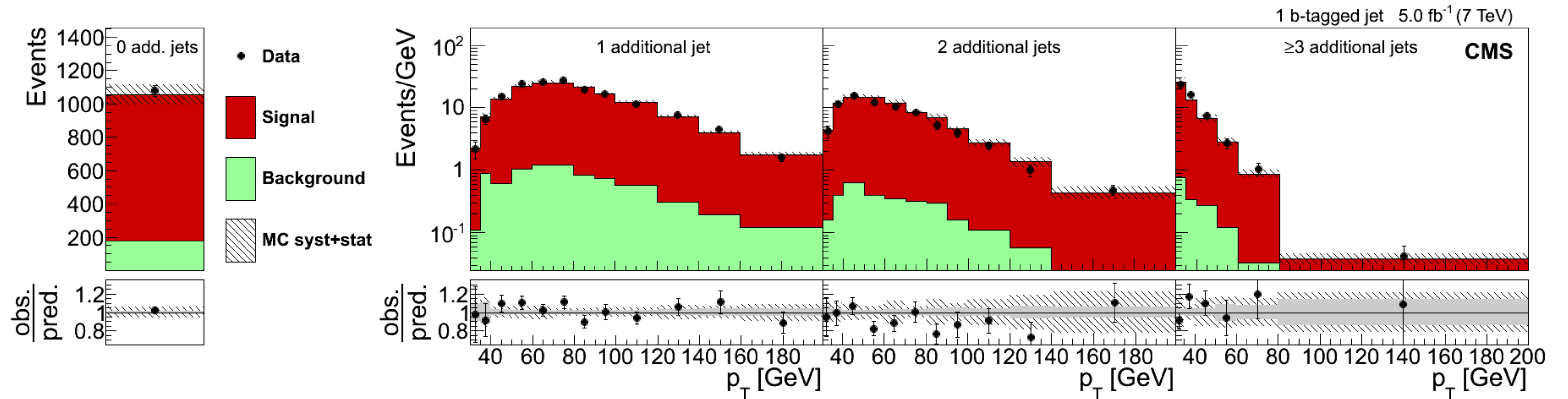}
\includegraphics[width=\textwidth]{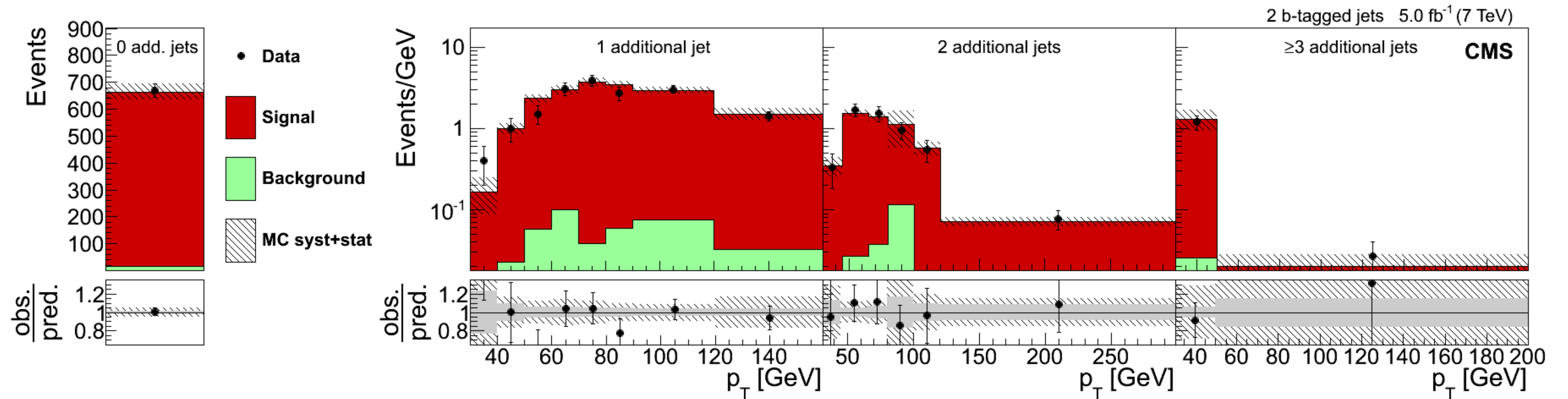}
\caption{Fitted total event yield for zero additional non-b-tagged jets (left)
and \pt of the non-b-tagged jet with the lowest \pt in the event (right) for events
with one, two, and at least three additional non-b-tagged jets, and with
zero or more than two (top row), one (middle row), and two (bottom row) b-tagged jets at $\sqrt{s} = 7\TeV$. The last bin of the \pt distributions includes the overflow events. The hatched bands correspond to the sum of statistical and systematic uncertainties in the event yield for the sum of signal and background predictions after the fit, and include all correlations. The ratios of data to the sum of the predicted yields are shown at the bottom of each plot. Here, an additional solid grey band represents the contribution from the statistical uncertainty in the MC simulation.
       \label{fig:lh_postfitdistr7}}
\end{figure}

\begin{figure}[htbp]
\centering
\includegraphics[width=\textwidth]{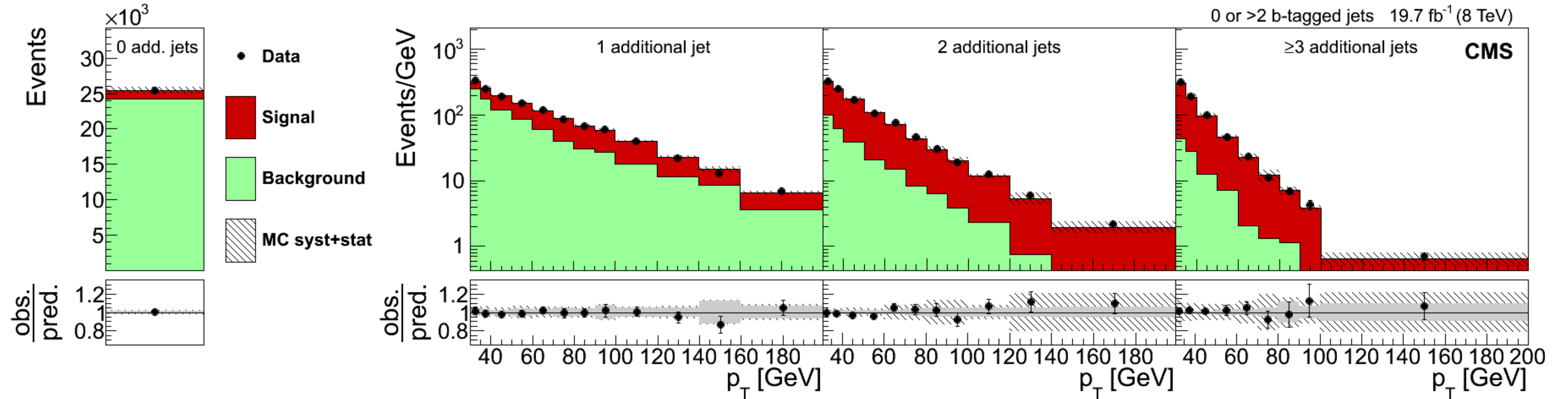}
\includegraphics[width=\textwidth]{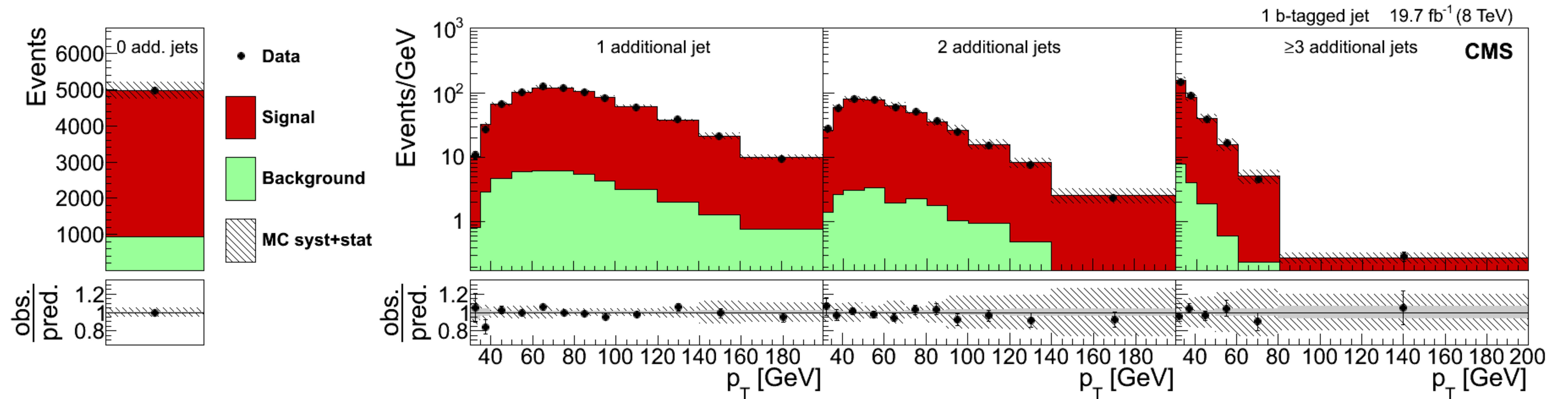}
\includegraphics[width=\textwidth]{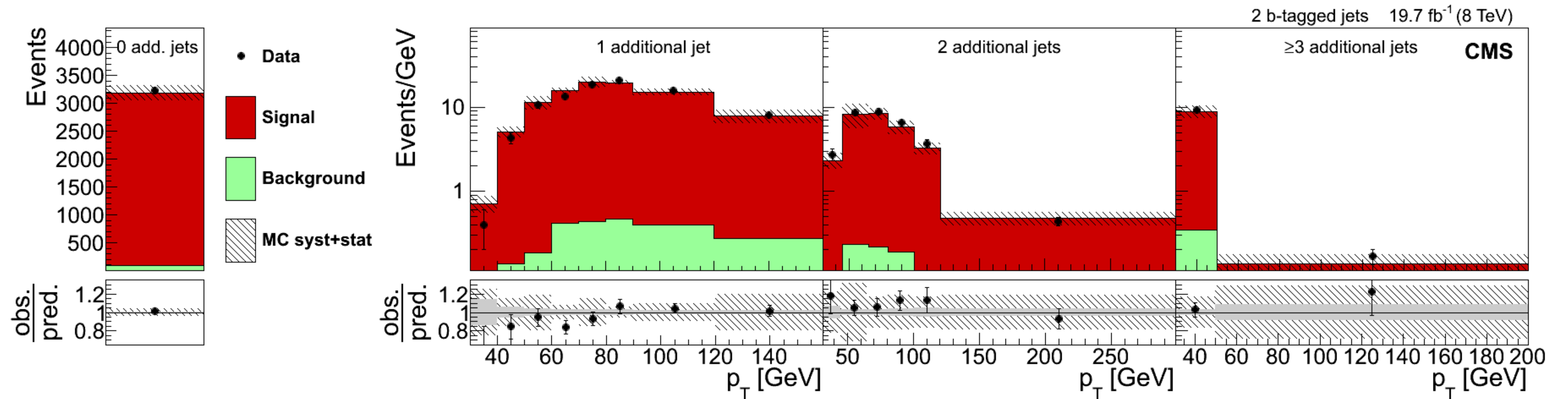}
\caption{Fitted total event yield for zero additional non-b-tagged jets (left)
and \pt of the non-b-tagged jet with the lowest \pt in the event (right) for events
with one, two, and at least three additional non-b-tagged jets, and with
zero or more than two (top row), one (middle row), and two (bottom row) b-tagged jets at $\sqrt{s} = 8\TeV$. The last bin of the \pt distributions includes the overflow events. The hatched bands correspond to the sum of statistical and systematic uncertainties in the event yield for the sum of signal and background predictions after the fit, and include all correlations. The ratios of data to the sum of the predicted yields are shown at the bottom of each plot. Here, an additional solid grey band represents the contribution from the statistical uncertainty in the MC simulation.
       \label{fig:lh_postfitdistr8}}
\end{figure}

Figures~\ref{fig:lh_postfitdistr7} and~\ref{fig:lh_postfitdistr8}
compare the data with the simulation after the simultaneous
fit at 7 and 8\TeV.
The uncertainty bands are calculated taking into account the full correlation matrix.
The description of the data by the simulation has improved with the fit. The best fit values of the nuisance parameters correspond to variations that are for most cases within $1\sigma$ of the prior uncertainties, about 98\% of the cases. The maximum observed variation is about $1.9\sigma$, corresponding to the uncertainty in the mistag SFs, see Section~\ref{sec:syst}. Other uncertainties with variations between 1 and 1.5$\sigma$ are two components of the jet energy scale corrections and the statistical component of the b tagging SFs.

The fiducial \ttbar production cross sections
at $\sqrt{s} = 7$ and 8\TeV are determined simultaneously.
For each centre-of-mass energy, a likelihood is defined as in Eq.~(\ref{eq:lhfunct}), respective
$\chi^2$ functions are constructed,
and the sum of both $\chi^2$ functions is minimised.
Correlations between systematic uncertainties are fully taken into account
(see Section~\ref{sec:syscorr}).

\subsection{Event counting  method}
\label{sec:xsecext_evc}
The \ttbar production cross section is also measured by applying an event counting method similar to the one used in a previous measurement~\cite{Chatrchyan:2013faa}. This method provides a cross-check of the reference method.

In this analysis, events are counted after applying the \emu selection described in Section~\ref{sec:eventsel} with additional requirements that help to further suppress the background contribution:
the presence of at least two jets is required, of which at least one has to be b-tagged. Compared with Ref.~\cite{Chatrchyan:2013faa}, tighter requirements on lepton isolation and identification, as well as on b tagging, are applied to further reduce the background contribution.

Techniques based on control samples in data are used to estimate the background contribution arising from DY and from non W/Z events. The contributions of the remaining background processes are estimated from simulation. The DY contribution is estimated using the ``$R_{\text{out/in}}$'' method~\cite{Chatrchyan:2013faa}, in which events with \eepm\ and \mmpm\ final states are used to obtain a normalisation factor.
This is estimated from the number of events within the $\PZ$ boson mass window in data, and extrapolated to the number of events outside the $\PZ$ mass window with corrections based
on control regions in data enriched in DY events. The contribution to the background originating from non $\PW$/$\PZ$ boson events is estimated by subtracting the same-sign prompt-lepton contributions from the same-sign event yields in data and multiplying by the ratio of opposite-sign over same-sign events. This ratio, originating from non-prompt lepton backgrounds, is taken from simulation.

Table~\ref{tab:yieldsChannels} shows the total number of events observed in
data and the numbers of expected signal and background events fulfilling all selection criteria.
For both data sets, a good agreement between data and expected number of events is observed.

\begin{table}[h]
\topcaption{Number of selected events for the event counting method for the 7 and 8\TeV data sets. The
results are given for the individual sources of background, \ttbar signal, and data.
The two uncertainties quoted correspond to the statistical and systematic components (cf. Section~\ref{sec:syst}), respectively.
}
\centering
\begin{tabular}{lcc}\hline
\multirow{2}{*}{Source}             & \multicolumn{2}{c}{Number of \emu events} \\
                                    & {7\TeV}             & {8\TeV}            \\\hline
DY                                  & $   22 \pm  3 \pm   3\phantom{0}  $  & $ 173 \pm 25 \pm   26\phantom{0}$   \\
Non W/Z                             & $   51 \pm  5 \pm  15  $  & $ 146 \pm 10 \pm   44\phantom{0}$   \\\hline
Single top quark ($\PQt \PW$)               & $  204 \pm  3 \pm  61\phantom{0}  $  & $1034 \pm  3 \pm  314\phantom{0}$   \\
${\rm \PV \PV}$                                  & $    7 \pm  1 \pm   2  $  & $  35 \pm  2 \pm   11$   \\
${\rm \ttbarV}$                             & $    12 \pm 1 \pm 3\phantom{0}  $  & $  84 \pm  1 \pm   26$   \\\hline
Total background                    & $  296 \pm 6 \pm  63\phantom{0}  $  & $ 1472 \pm 27 \pm  319\phantom{0}$   \\\hline
$\mathrm{t\bar{t}}$ dilepton signal & $ 5008 \pm 15 \pm 188\phantom{0}  $  & $24440 \pm 44 \pm  956\phantom{00}$   \\\hline
Data                                & $ 4970                       $  & $25441$   \\
\hline
\end{tabular}
\label{tab:yieldsChannels}
\end{table}

\begin{figure}[htbp!]
\centering
\includegraphics[width=0.495\textwidth]{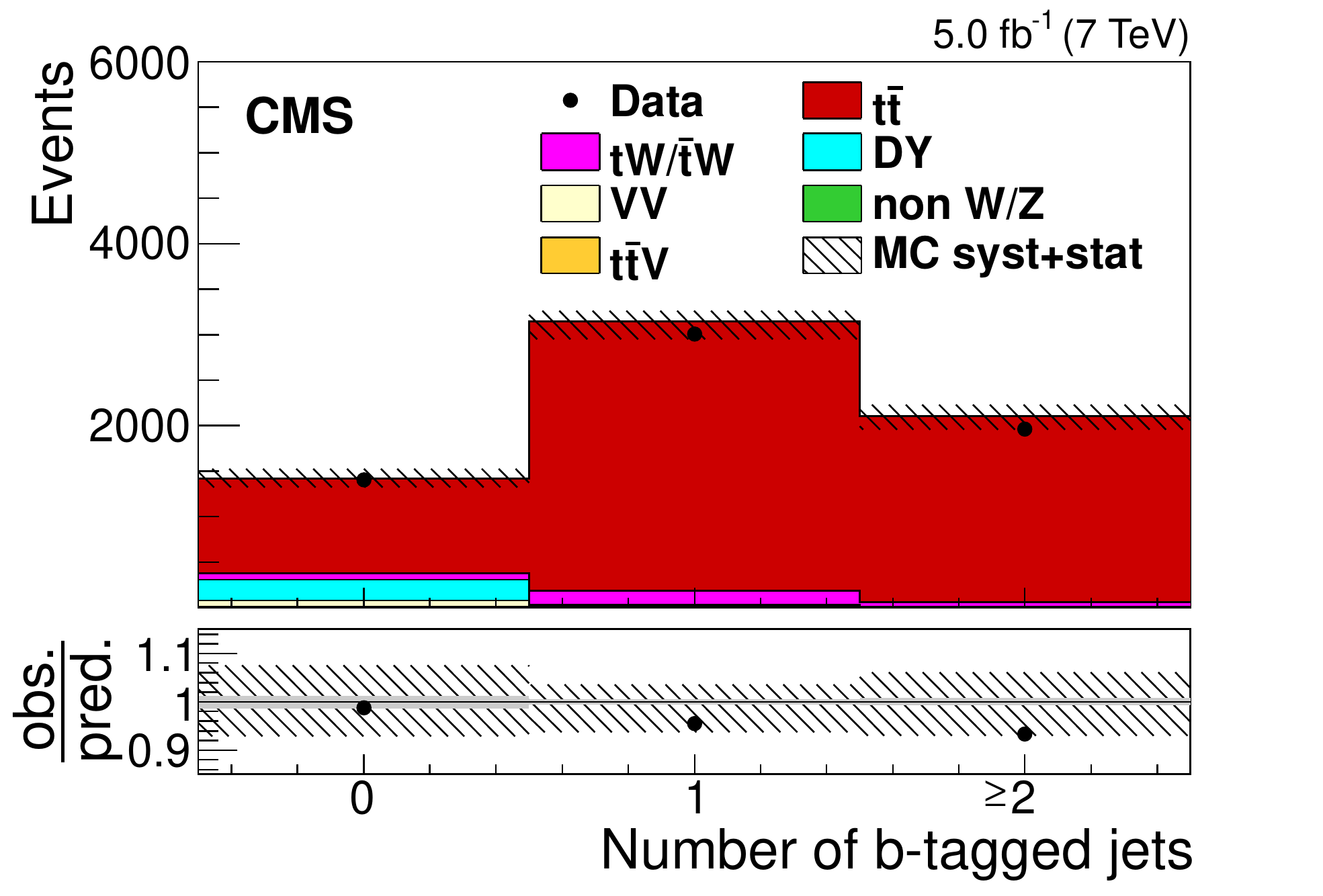}
\includegraphics[width=0.495\textwidth]{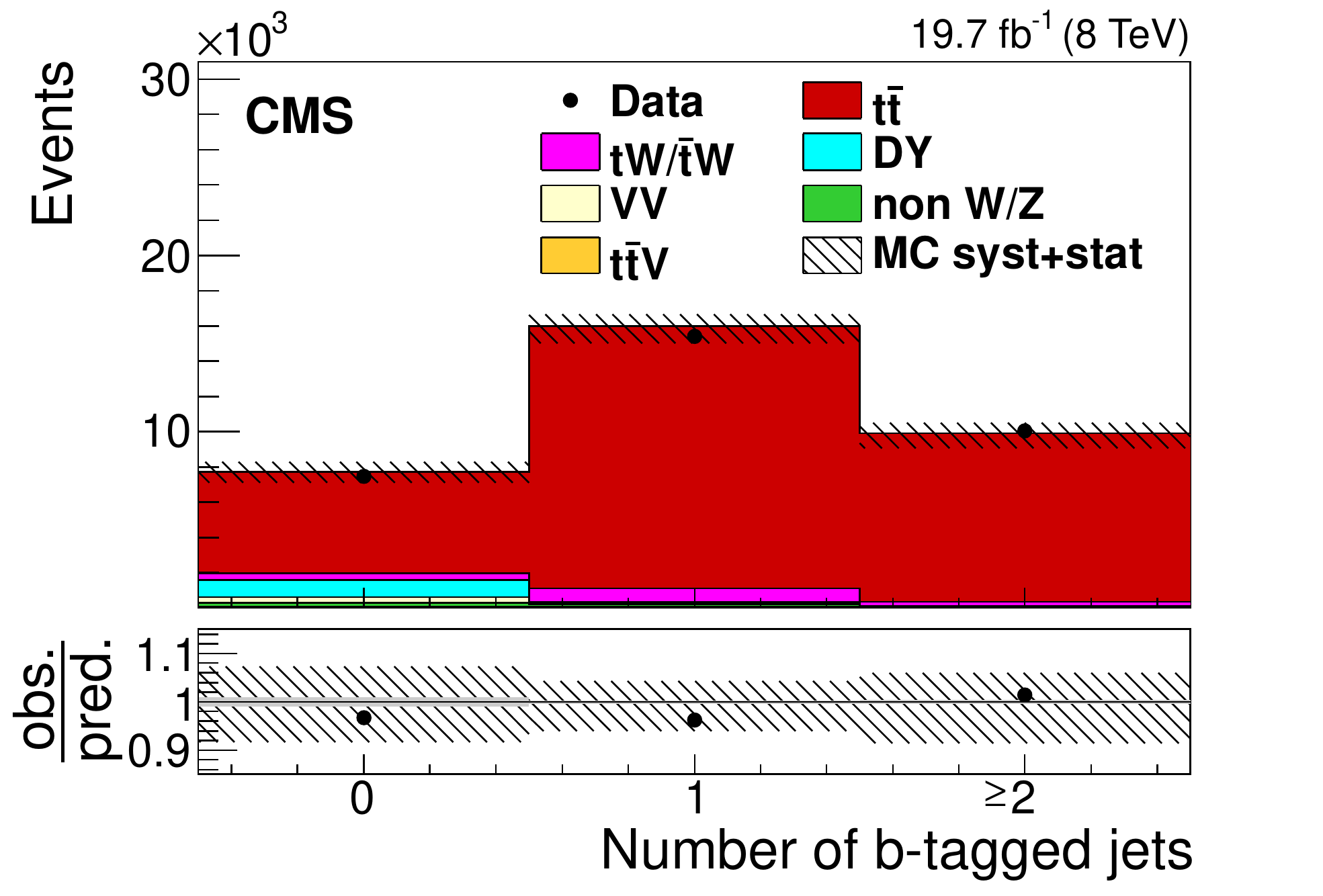}
\caption{
Comparison of the b jet multiplicity distributions in the \emu channel for 7 (left) and 8 (right)\TeV between the data and simulation for events fulfilling the \emu selection and the requirement of having at least two jets. The hatched bands correspond to the sum of statistical and systematic uncertainties in the event yield for the signal and background predictions. The ratios of data to the predicted yields are shown at the bottom of each plot. Here, an additional solid grey band represents the contribution from the statistical uncertainty in the MC simulation.
\label{fig:nbjets_cc}}
\end{figure}

Figure~\ref{fig:nbjets_cc} shows the b jet multiplicity in events passing the full event selection,
except for the b jet requirement, for data collected at 7 and 8\TeV. In both cases the total predicted
yields provide a good description of the measured distributions.

The cross section $\stt$ is determined from the number of data events after background subtraction, and dividing by the integrated luminosity of the data sample and by the product of detector and kinematical acceptance, selection efficiency, as estimated from simulation for a top quark mass of 172.5\GeV, and branching fraction of the
selected \ttbar dilepton final state.

\section{Systematic uncertainties}
\label{sec:syst}
The measurement of the top quark pair production cross section is affected by systematic uncertainties that originate
from detector effects and from theoretical assumptions.
Each source of
systematic uncertainty is assessed individually
by suitable variations of the MC simulations or by varying
parameter values within their estimated uncertainties in the analysis.
Each source is represented by
a nuisance parameter, which is fitted together with
$\sigma^{\text{vis}}_{\ttbar}$, as described in Section~\ref{sec:xsecext}.
For the event counting method, the same sources of systematic uncertainty
are evaluated following the procedure in Ref.~\cite{Chatrchyan:2013faa}.

\subsection{Experimental uncertainties}
\label{sec:syst_exp}

The uncertainty in the dilepton trigger (``Trigger'')
and lepton identification efficiencies (``Lepton ID/isolation'')
are estimated by varying the SFs within their uncertainties,
which are in the range of 1--2\%.

The lepton energies (``Lepton energy scale'') are corrected separately for electrons~\cite{Khachatryan:2015hwa} and for muons~\cite{Bodek:1473782}. Their scales are varied by 0.15\% for electrons and 0.3\% for muons.

The uncertainty due to the limited knowledge of the jet energy scale (``JES'') is determined by variations of the jet energy in bins of \pt and $\eta$~\cite{bib:JME-10-011:JES}.
For the reference method, these variations are divided into 27 sources and the effect of each source is evaluated individually. For the event counting method, the total variation is used to determine the uncertainty.

The uncertainty due to the limited accuracy of the jet energy resolution (``JER'')
is determined by changing the simulated JER by $\pm2.5\%$, $\pm4\%$, and $\pm5\%$, for jets with $|\eta| < 1.7$, $1.7 < |\eta| < 2.3$, and $
|\eta| > 2.3$, respectively~\cite{bib:JME-10-011:JES}.

For the normalisation of each background source, an uncertainty of $\pm$30\%
is assumed. In the case of the single top quark background (``$\PQt \PW$/$\PAQt \PW$''), the variation covers the uncertainty in the absolute rate, including uncertainties due to PDFs.
The same global variation is applied to the other dominant background contribution, DY.
The predicted cross section has an uncertainty of $\approx$5\%, including PDF uncertainties.
The variation used here additionally covers the observed differences in heavy-flavour composition between data and simulation in dedicated CMS analyses and is also suggested by estimates based on data~\cite{CMStopPublication2,top12001}.

The uncertainties due to the b tagging efficiency (``b tag'')
and misidentification rate (``Mistag'') are determined
by varying the b tagging SFs of the b jets or the light-flavour jets, respectively, by the uncertainties quoted in Ref.~\cite{BTV-11-004-pub}.
For the reference method, the b tagging uncertainties are divided into 19 individual sources, some of them are correlated to other systematic uncertainties, such as JER or pileup. The remaining sources are evaluated individually.

The effect of pileup events (``Pileup'') is evaluated by weighting the inelastic pp cross section in simulation to the minimum bias cross section determined in data. The pileup model estimates the mean number of additional pp interactions to be about 9 events for the data collected at 7\TeV and 21 for the data collected at 8\TeV. These estimates are based on the total inelastic pp cross sections at $\sqrt{s}=$~7~(8)\TeV, which are determined to be 73.5 (69.4)\unit{mb}, following the measurement described in Ref.~\cite{bib:ppInelXSec}. The systematic uncertainty is determined by varying the cross sections within their uncertainty, $\pm$8\% at 7\TeV and $\pm$5\% at 8\TeV.

The uncertainty in the luminosity (``Luminosity'') measurement is 2.2\%~\cite{CMS-PAS-SMP-12-008} at 7\TeV and 2.6\%~\cite{CMS-PAS-LUM-13-001} at 8\TeV.

\subsection{Theoretical uncertainties}
\label{sec:syst_model}

The impact of theoretical assumptions in the modelling is determined
by repeating the analysis and replacing the standard \MADGRAPH $\ttbar$ simulation
by dedicated simulation samples with varied parameters.

The uncertainty in modelling of the hard-production process
(``$Q^2$ scale'') is assessed through a simultaneous variation of renormalisation and factorisation scales in the \MADGRAPH sample by factors of 2 and 0.5
relative to their common nominal value, which is set to the $\mu_\mathrm{F}^2=\mu_\mathrm{R}^2=Q^2$ scale of the hard process.
In \MADGRAPH, it is defined by $Q^2 = m^2_{\PQt} + \Sigma \pt^2$,
where the sum is over all additional final state partons in the matrix element calculations.

The impact of the choice of the scale
that separates the description of jet production through matrix elements or parton shower
(``ME/PS matching'')
in \MADGRAPH is studied by changing its reference value of 20\GeV to 40\GeV and to 10\GeV.

The effect of the matrix-element generator choice on the measurement is evaluated by using \POWHEG~\cite{bib:powheg1,bib:powheg2,bib:powheg3} for the \ttbar simulation instead of \MADGRAPH (``\MADGRAPH vs \POWHEG'').

The flavour-dependent hadronisation uncertainty (``Hadronisation (JES)'') is part of the JES uncertainty and comes from differences in the jet energy response for different jet flavours.
It is estimated by the differences between using simulations
with the Lund fragmentation model in \PYTHIA and cluster fragmentation model in \HERWIGpp~\cite{HERWIGPP} and is evaluated for each jet flavour independently.
An additional uncertainty included in this source is
the uncertainty in the b quark fragmentation tune.
This is  evaluated by varying the Bowler--Lund $\PQb$ quark fragmentation
model in tune Z2* to describe the results by ALEPH~\cite{Heister:2001jg} and DELPHI~\cite{DELPHI:2011aa} for the $\PQb$ quark fragmentation functions. Another uncertainty included in this source is the uncertainty in the semileptonic branching fraction of B hadrons, varied between $10.05$\% and $11.27$\%, which is the range of the measurements from $\PBz$/$\PBp$ decays and their uncertainties~\cite{Agashe:2014kda}.

Differential cross section measurements~\cite{bib:TOP-12-028}
have shown that the \pt of the top quark is softer
than predicted by the nominal \MADGRAPH simulation used to measure the cross section.
To account for this effect, the difference between the result obtained
with the nominal simulation and using the \MADGRAPH prediction reweighted to describe the measured top quark \pt spectrum is taken as a systematic uncertainty (``Top quark \pt modelling'').

The uncertainties from ambiguities in modelling colour reconnection effects (``Colour reconnection'')
are estimated by comparing simulations of an underlying event tune including colour reconnection
to a tune without it, the Perugia 2011 (P11) and P11 noCR tunes~\cite{Skands:2010ak}.

The uncertainty in the modelling of the underlying event (``Underlying event'') is estimated by evaluating the relative variations of two different
P11 \PYTHIA tunes with respect to the standard P11 tune: the mpiHi and the TeV tunes with higher and lower underlying event activity, respectively.

The uncertainty from the choice of PDFs (``PDF'') is determined
by reweighting the sample of simulated \ttbar events according
to the 52 CT10 error PDF sets~\cite{bib:CT10}, scaled to 68\% CL.

\subsection{Correlations between systematic uncertainties for the measurements at 7 and \texorpdfstring{8\TeV}{8TeV}}
\label{sec:syscorr}
A number of systematic uncertainties affect the measurements at $\sqrt{s} = 7$ and 8\TeV similarly,
while others are completely decoupled. In this analysis, systematic uncertainties are treated as
either uncorrelated, partially correlated, or fully correlated
between the two measurements.
For fully correlated systematic uncertainties, common nuisance
parameters are used in the simultaneous likelihood
fit to the two data sets.
For each partially correlated systematic uncertainty source, three nuisance
parameters are introduced, one for each data set for the uncorrelated
part and one common parameter for the correlated part.
The degree of correlation is modelled by the parameter $\rho$.
The uncertainties of the correlated and the two uncorrelated parameters are taken to
be fractions $\rho$ and $\sqrt{\smash[b]{1-\rho^2}}$, respectively,
of the uncertainty of the original nuisance parameter.
The $\rho$ values assumed for this analysis are listed in Table~\ref{tab:corr_priors}.

\begin{table}[h]
\topcaption{Assumed correlations $\rho$
between systematic uncertainties for the 7 and 8\TeV data sets.
If $\rho=0$, the uncertainties are treated as uncorrelated
between the two sets.
}
 \centering
        \begin{tabular}{lc}
        \hline
         Uncertainty source & $\rho$ \\ \hline
         Trigger & 0.8 \\
         Electron ID & 0.9 \\
         Electron energy scale & 0.9 \\
         Muon ID & 0.9 \\
         Muon energy scale & 0.9 \\
         JES: flavour & 1 \\
         JES: pileup & 0 \\
         JES: absolute extrapolation & 1 \\
         JES: other & 0 \\
         Jet energy resolution & 0.9 \\
         Each background & 0.9 \\
         b-tag (JES) & 0.2 \\
         b-tag (stat) & 0 \\
         b-tag (syst) & 1 \\
         Mistag & 0.8 \\
         Pileup & 0.5 \\
         \hline
         $\mu_\mathrm{R}$, $\mu_\mathrm{F}$ scales & 1 \\
         ME/PS matching & 1 \\
         \MADGRAPH vs \POWHEG & 1 \\
         b quark fragmentation tune & 1 \\
         B hadron semileptonic branching fraction & 1 \\
         Top quark \pt modelling & 1 \\
         Colour reconnection & 1 \\
         Underlying event & 1 \\
         PDF & 1 \\
         \hline
         Integrated luminosity & 0 \\ \hline
         \end{tabular}
 \label{tab:corr_priors}
\end{table}

For experimental sources, the same procedures
are usually employed at the two centre-of-mass energies
for calibration and determination of uncertainties. Also, the same MC generators are used for the modelling of background processes. Hence, these uncertainties are treated as 100\% correlated,
however for each source a (usually small) uncorrelated component arises
from statistical fluctuations in the data or simulated samples. The resulting correlation coefficients are estimated to be 0.9
for several sources and 0.8 for the ``Trigger'' and ``Mistag'' sources. For the ``Pileup'' source a relatively small correlation
of 0.5 is assumed because of the largely different beam conditions
at the two energies.

From the uncertainties related to the JES, the flavour components (``JES: flavour''), owing to the comparison between different hadronisation models, and components related to the extrapolation from $\PZ \to \ell\ell$ kinematic acceptance to the full phase space using MC simulation (``JES: absolute extrapolation'') are taken as fully correlated. The JES sources related to pileup (``JES: pileup'') are treated as uncorrelated, because of different procedures used for the uncertainty assessment at the two energies, as well as the remaining terms (``JES: other''). The JES component of the b tagging uncertainties is fitted independently, assigning a correlation coefficient of 0.2 that reflects the amount of correlated JES uncertainty sources.

All modelling uncertainties are assumed to be fully correlated
between the two centre-of-mass energies, including the three remaining JES parts.
The integrated luminosity uncertainties are treated as fully uncorrelated, to account for the different beam conditions and specific effects associated to each measurement.
It has been checked that variations of the assumed correlations
within reasonable ranges lead to negligible changes of the extracted
cross sections.
\subsection{Final uncertainties}
The total uncertainties in the fiducial cross sections, as obtained with the binned likelihood fit (Section~\ref{sec:xsect_binlh}), are
$^{+3.6}_{-3.4}\,$\% at 7\TeV and $^{+3.7}_{-3.4}\,$\% at 8\TeV.
The impact of the sources of systematic uncertainties in this total uncertainty are listed in Table~\ref{tab:lh_syst_sum}. These are estimated by removing groups of uncertainties one at a time and gauging the difference in quadrature on the total uncertainty. Significant contributions to the total uncertainty spread over many different sources of experimental and modelling uncertainties with
``Lumi, '', ``Lepton ID/isolation'', ``Trigger'', and ``DY'' being
the four largest sources. The observed shifts of the fitted background or other nuisance parameters
compared to their assumed uncertainty before the fit are in general
small, indicating a consistent fit.

\begin{table}[h]
\topcaption{
Illustrative summary of the individual contributions to the total uncertainty in
the visible \ttbar cross section measurements.
}
\center
\begin{tabular}{lcc}
\hline
\multirow{2}{*}{Source}     &  \multicolumn{2}{c}{Uncertainty [\%]}  \\
                            &  7\TeV  &  8\TeV  \\
\hline
Trigger                     &  ${1.3}$ & ${1.2}$ \\
Lepton ID/isolation         &  ${1.5}$ & ${1.5}$ \\
Lepton energy scale         &  ${0.2}$ & ${0.1}$ \\
Jet energy scale            &  ${0.8}$ & ${0.9}$\\
Jet energy resolution       &  ${0.1}$   & ${0.1}$ \\
$\PQt \PW$/$\PAQt \PW$                    &  ${1.0}$ & ${0.6}$ \\
DY                          &  ${1.4}$ & ${1.3}$ \\
\ttbg.             &  ${0.1}$ & ${0.1}$ \\
${\rm \ttbarV}$      &  ${0.1}$   & ${0.1}$ \\
Diboson           &  ${0.2}$ & ${0.6}$ \\
W+jets/QCD        &  ${0.1}$   & ${0.2}$ \\
b-tag                       &  ${0.5}$ &  ${0.5}$  \\
Mistag                      &  ${0.2}$ &  ${0.1}$ \\
Pileup                     &  ${0.3}$ &  ${0.3}$ \\
\hline
$\mu_\mathrm{R}$, $\mu_\mathrm{F}$ scales                &  ${0.3}$ &  ${0.6}$ \\
ME/PS matching              &  ${0.1}$ &  ${0.1}$ \\
\MADGRAPH vs \POWHEG   &  ${0.4}$ &  ${0.5}$ \\
Hadronisation (JES)         &  ${0.7}$ &  ${0.7}$ \\
Top quark \pt modelling                &  ${0.3}$ &  ${0.4}$ \\
Colour reconnection          &  ${0.1}$ &  ${0.2}$ \\
Underlying event            &  ${0.1}$   &  ${0.1}$ \\
PDF                         &  ${0.2}$ &  ${0.3}$ \\
\hline
Integrated luminosity                  &  ${2.2}$  &  ${2.6}$\\
\hline
Statistical                 &  ${1.2}$ & ${0.6}$ \\
\hline
\end{tabular}
\label{tab:lh_syst_sum}
\end{table}

\clearpage

\section{Cross section measurement}
\label{sec:results}
The results of the \ttbar cross section measurements in pp collisions at~7 and 8\TeV
are presented in the fiducial range
and in the full phase space.

\subsection{Fiducial cross section}
\label{sec:xsec_fiduc}
The fiducial cross sections are defined
for $\ttbar$ production with events containing an oppositely charged \emu pair with both leptons having $\pt > 20\GeV$ and $\abs{\eta} < 2.4$.
The measured cross sections, using the binned likelihood fit extraction method
(Section~\ref{sec:xsecext}) and assuming a top quark mass of $172.5\GeV$, are
\begin{equation*}\begin{aligned}
 \sigma^{\text{vis}}_{\ttbar} & = 
 3.03 \pm 0.04\stat ^{+\,0.08}_{-\,0.07}\syst\pm 0.07\lum\unit{pb},
\quad
\text{at $\sqrt{s}=7$\TeV and}
\\
\sigma^{\text{vis}}_{\ttbar} & = 
 4.23 \pm 0.02\stat ^{+\,0.11}_{-\,0.09}\syst\pm 0.11\lum\unit{pb},
\quad \text{at  $\sqrt{s}=8$\TeV.}
\end{aligned}\end{equation*}
The uncertainties are due to statistical fluctuations,
combined experimental and theoretical systematic effects
on the measurement, and the uncertainty in the measurement of the integrated luminosity.
A summary of the systematic uncertainties is presented in Table~\ref{tab:lh_syst_sum}.

\subsection{Full phase space cross section}
\label{sec:xsec_tot}
The full phase space (total) cross sections for \ttbar production are calculated from the fiducial cross section results by dividing $\sigma^{\text{vis}}_{\ttbar}$
by the acceptance, as in Eq.~(\ref{eq:extrapol}).
The quantity $A_{\emu}$ is determined from the
$\ttbar$ signal MC simulation.
As it depends on the exact theoretical model used in the
event generation part of the simulation, it is parametrised
as a function of the same nuisance parameters that
were used for the modelling uncertainties (Section~\ref{sec:syst})
in the binned likelihood fit extraction of the fiducial cross sections.
The fitted values of these nuisance parameters are used to obtain the best estimates
of $A_{\emu}$, $1.745\times 10^{-2}$ at 7\TeV and $1.728\times 10^{-2}$ at 8\TeV, which are used for the determination of the nominal values of \stt.
In order to determine the uncertainty in the phase space
extrapolation modelled by $A_{\emu}$,
each relevant nuisance parameter is iteratively varied from the fitted value
by the ${\pm}1\sigma$ values before the fit, while all other
nuisance parameters are kept at their fitted values. The resulting variations
of $A_{\emu}$ are taken as an additional extrapolation uncertainty.
The sources that are considered here are
``$\mu_\mathrm{R}$ and $\mu_\mathrm{F}$ scales'', ``ME/PS matching'', ``Top quark \pt modelling'', and ``PDF'' (see Section~\ref{sec:syst}),
and the individual uncertainties in $\stt$ from these sources are
added in quadrature. The resulting systematic uncertainties are listed in Table~\ref{tab:lh_syst_sumex}.

\begin{table}[h]
\topcaption{Individual contributions to the systematic uncertainty in
the total \ttbar cross section measurements.
The total systematic uncertainties in the fiducial cross sections $\sigma^{\text{vis}}_{\ttbar}$ are given in the row ``Total (visible)'', and those in the full phase space cross section $\stt$  in the row  ``Total''.
}
\center
\renewcommand{\arraystretch}{1.2}
\begin{tabular}{lcc}
\hline
\multirow{2}{*}{Source}     &  \multicolumn{2}{c}{Uncertainty [\%]}  \\
                            &  7\TeV  &  8\TeV  \\
\hline
Total  (visible)                 & $^{+3.6}_{-3.4}$  &  $^{+3.7}_{-3.4}$ \\ \hline
$Q^{2}$ scale (extrapol.)    &  $^{+0.1}_{-0.4}$   & $^{+0.2}_{-0.1}$ \\
ME/PS matching (extrapol.)   & $^{+0.1}_{-0.1}$   & $^{+0.3}_{-0.3}$ \\
Top quark \pt (extrapol.)      &  $^{+0.5}_{-0.3}$   & $^{+0.6}_{-0.3}$ \\
PDF (extrapol.)              &  $^{+0.1}_{-0.1}$   & $^{+0.1}_{-0.1}$ \\
\hline
Total                        &  $^{+3.6}_{-3.5}$  &  $^{+3.7}_{-3.5}$ \\
\hline
\end{tabular}
\renewcommand{\arraystretch}{1.0}

\label{tab:lh_syst_sumex}
\end{table}

The measurements of \stt at the two centre-of-mass energies are
\begin{equation*}\begin{aligned}
\sigma_{\ttbar} & =
 173.6 \pm 2.1\stat ^{+\,4.5}_{-\,4.0}\syst\pm 3.8\lum\unit{pb},
\quad
\text{at $\sqrt{s}=7$\TeV and}
\\
\sigma_{\ttbar} & =
 244.9 \pm 1.4\stat ^{+\,6.3}_{-\,5.5}\syst\pm 6.4\lum\unit{pb},
\quad \text{at  $\sqrt{s}=8$\TeV.}
\end{aligned}\end{equation*}
After adding the uncertainties in quadrature,
the resulting total uncertainties are
$6.2\unit{pb}$\,(3.6\%) at $\sqrt{s}=7$\TeV
and
$9.1\unit{pb}$\,(3.7\%) at $\sqrt{s}=8$\TeV.

The results obtained with the method based on
event counting (see Section~\ref{sec:xsecext_evc}) are
\begin{equation*}\begin{aligned}
\sigma_{\ttbar} & =
165.9 \pm 2.5\stat \pm \phantom{0}6.2\syst\pm 3.6\lum\unit{pb},
\quad
\text{at $\sqrt{s}=7$\TeV and}
\\
\sigma_{\ttbar} & =
 241.1 \pm 1.6\stat \pm 10.0\syst\pm 6.3\lum \unit{pb},
\quad \text{at  $\sqrt{s}=8$\TeV.}
\end{aligned}\end{equation*}
As expected, the statistical and systematic uncertainties
are slightly larger than those obtained with the reference method.
The results of the two methods are in agreement.

The cross section measurements agree with previous results~\cite{CMStopPublication2,top12001,Chatrchyan:2013faa,
ATLAStopPublication5,Aad:2014kva,ATLAStopPublication2,ATLAStopPublication1}.
They constitute the most precise CMS measurements of \stt\
to date and have a similar precision to the most precise ATLAS result~\cite{Aad:2014kva}, obtained in
the same decay channel.
For both centre-of-mass energies,
the predicted cross sections at NNLO (see Section~\ref{sec:datasim})
are in good agreement with the measurements.

The ratio of cross sections using the results obtained with the reference analysis amounts to
\begin{equation*}
R_{\ttbar} = \sigma_{\ttbar}(8\TeV)/\sigma_{\ttbar}(7\TeV)= 1.41 \pm 0.06.
\end{equation*}
Here, the correlated uncertainty obtained from the simultaneous likelihood fit (Section~\ref{sec:xsecext})
of the fiducial cross sections at the two centre-of-mass energies is fully taken into account as
well as the correlated uncertainty on the acceptances arising from model uncertainties, which are assumed to be fully correlated between
the two energies. The total relative uncertainty of the ratio is $4.2\%$,
indicating a partial cancellation of systematic uncertainties. The predicted ratio at NNLO (see Section~\ref{sec:datasim}) is consistent with the measurement.

\section{Determination of the top quark pole mass}
\label{sec:mtpole}
The full phase space cross sections
are used to determine the top quark pole mass ($m_{\PQt}$) via the dependence
of the theoretically predicted cross section on $m_{\PQt}$
and comparing it to the measured cross section. For this purpose,
the cross section fit and the extrapolation to the full phase space (see Sections~\ref{sec:xsecext} and~\ref{sec:xsec_tot})
are repeated for three different hypotheses for the top quark mass parameter in the MC simulation ($m_{\PQt}^{\mathrm{MC}}$): 169.5, 172.5, and 175.5 \GeV. For each mass value a sample of simulated \ttbar events, generated with the corresponding $m_{\PQt}^{\mathrm{MC}}$ value, is used
in the fit as a signal model. The dependence of the distributions used in the fit on detector effects is evaluated individually for each mass value. Their dependence on modelling uncertainties
varies little over the studied mass range and
is thus taken from the nominal mass value ($m_{\PQt}^{\mathrm{MC}}$ = 172.5\GeV). The obtained cross section dependence on the mass can be parametrised as an exponential function:
\begin{equation*}
\begin{aligned}
\sigma_\ttbar(7\TeV,m_{\PQt}^{\mathrm{MC}}) & =  \exp {\left[-0.1718 \, (m_{\PQt}^{\mathrm{MC}}/\GeVns  -178.5 ) \right]} + 170.9\unit{pb}, \\
\sigma_\ttbar(8\TeV,m_{\PQt}^{\mathrm{MC}}) & =  \exp {\left[-0.1603 \, (m_{\PQt}^{\mathrm{MC}}/\GeVns  -185.4 ) \right]} + 237.0\unit{pb}.
\end{aligned}\end{equation*}

To express the measured dependence as a function of $m_{\PQt}$ instead of $m_{\PQt}^{\mathrm{MC}}$, the  difference between $m_{\PQt}$ and $m_{\PQt}^{\mathrm{MC}}$ needs to be accounted for. This is estimated to be of the order of 1\GeV~\cite{Buckley:2011ms}. Therefore, an additional uncertainty $\Delta_{m_{\PQt}\pm}$ in the obtained cross section dependence is introduced. It is evaluated by shifting the measured dependence by $\pm1\GeV$ in $m_{\PQt}^{\mathrm{MC}}$ and recording the difference in $\sigma_{\ttbar}$.
For the determination of $m_{\PQt}$, this contribution to the total uncertainty is almost negligible. In consequence, the measurements of \stt\ can be represented by Gaussian likelihoods as a function of $m_{\PQt}$ of the form
\begin{equation}
L_\text{exp}(m_{\PQt},\sigma_{\ttbar}) = \exp \left[\frac{\left(\sigma_{\ttbar}(m_{\PQt}) - \sigma_{\ttbar}\right)^2}{-2(\Delta^2  +  \Delta_{m_{\PQt}\pm}^2)}\right],
\end{equation}
where $\Delta$ represents the total uncertainty in each of the cross section measurements and $\sigma_{\ttbar}(m_{\PQt})$ the measured dependence of the cross section on  $m_{\PQt}$.

The predicted dependence of $\sigma_{\ttbar}$ on the top quark pole mass at NNLO+NNLL is determined with \textsc{TOP++}, employing different PDF sets (NNPDF3.0~\cite{Ball:2014uwa}, CT14~\cite{Dulat:2015mca}, and MMHT2014~\cite{Harland-Lang:2014zoa}) with $\alpha_\mathrm{s}=0.118\pm 0.001$. Additionally, uncertainties of 1.79\% at 7\TeV and 1.72\% at 8\TeV are assigned to the predicted cross section values to account for the uncertainty in the LHC beam energy~\cite{Wenninger:1546734}. The predicted \stt\ is represented by an asymmetric Gaussian function with width $\Delta_{\mathrm{p},\pm}$, comprising PDF, $\alpha_\mathrm{s}$, and the beam energy uncertainty summed in quadrature. This function is convolved with a box function to account for the uncertainty in the renormalisation and factorisation scales in the prediction~\cite{top_alphas}. The result of the convolution is given as
\begin{equation}
L_{\text{pred}}(m_{\PQt},\sigma_{\ttbar}) = \frac{1}{C(m_{\PQt})}
\left( \text{erf} \left[ \frac{\sigma_{\ttbar}^{({\rm h})}(m_{\PQt})-\sigma_{\ttbar}}{\sqrt{2} \Delta_{\mathrm{p},+}} \right]
- \text{erf} \left[ \frac{\sigma_{\ttbar}^{({\rm l})}(m_{\PQt})-\sigma_{\ttbar}}{\sqrt{2} \Delta_{\mathrm{p},-}} \right]
\right) \text{,}
\end{equation}
where $\sigma_{\ttbar}^{({\rm h})}$ and $\sigma_{\ttbar}^{({\rm l})}$ denote the upper and lower predicted cross section values, respectively, from variations of the renormalisation and factorisation scales. The normalisation factor $C(m_{\PQt})$ assures that $\max({L_{\text{pred}}})=1$ for any fixed $m_{\PQt}$.

\begin{figure}[htbp]
\centering
\includegraphics[width=0.65\textwidth]{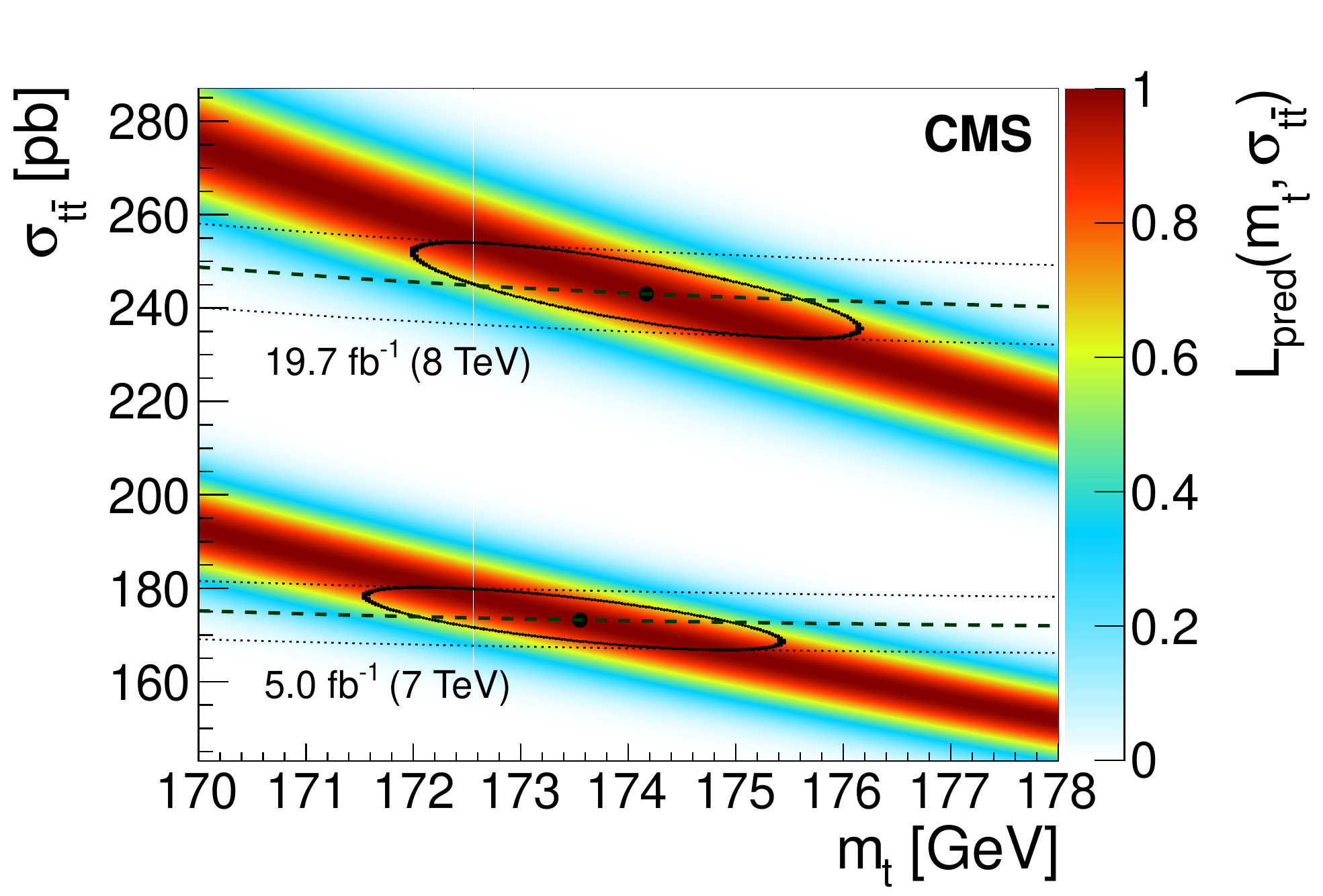}
\caption{Likelihood for the predicted dependence of the \ttbar production cross section on the top quark pole mass for 7 and 8\TeV determined with \textsc{TOP++}, employing the NNPDF3.0 PDF set. The measured dependences on the mass are given by the dashed lines, their 1$\sigma$-uncertainties are represented by the dotted lines. The extracted mass at each value of $\sqrt{s}$ is indicated by a black point, with its 1$\sigma$-uncertainty constructed from the continuous contour, corresponding to $-2\Delta \log (L_\text{pred} L_\text{exp}) = 1$.
\label{fig:topmass_post8}}
\end{figure}

Figure~\ref{fig:topmass_post8} shows the likelihoods for the predicted \ttbar cross section employing NNPDF3.0 and the measurement of \stt\ at $\sqrt{s}=7$ and 8\TeV as a function of $m_{\PQt}$. The product of the two likelihoods is used to fit the mass value by maximizing
the likelihood simultaneously with respect to $m_{\PQt}$ and $\sigma_{\ttbar}$. The extracted top quark pole masses using different PDF sets are listed in Table~\ref{tab:mtpoleenergies}. The contributions from uncertainties in the CT14 PDF set are scaled to a 68\% CL.

\begin{table}[h]
\topcaption{Top quark pole mass at NNLO+NNLL extracted by comparing the measured \ttbar production  cross section at 7 and 8\TeV with predictions employing different PDF sets. }
\centering
\renewcommand{\arraystretch}{1.25}
        \begin{tabular}{ l  c   c }
                \hline
     \multirow{2}{*}{}     &  \multicolumn{2}{c}{$m_{\PQt}$ [\GeV]}  \\
                            &  7\TeV  &  8\TeV  \\
        \hline
NNPDF3.0   &   $173.5^{+1.9}_{-2.0}$  &   $174.2^{+2.0}_{-2.2}$  \\
MMHT2014  &   $173.9^{+2.0}_{-2.1}$  &   $174.4^{+2.1}_{-2.3}$   \\
CT14      &   $174.1^{+2.2}_{-2.4}$  &   $174.6^{+2.3}_{-2.5}$ \\
\hline
\end{tabular}

 \label{tab:mtpoleenergies}
\end{table}

Finally, a weighted average is calculated,
taking into account all systematic uncertainty correlations between
the measured cross sections at 7 and 8\TeV,
and assuming 100\% correlated uncertainties for the theoretical predictions
at the two energies.
The resulting top quark pole masses are listed in Table~\ref{tab:mtpolepdfs} and are in good agreement with each other and previous measurements~\cite{Aad:2014kva,top_alphas}.

\begin{table}[h]
\topcaption{Combined top quark pole mass at NNLO+NNLL extracted by comparing the measured \ttbar production cross section with predictions employing different PDF sets.}
\centering
\renewcommand{\arraystretch}{1.25}
        \begin{tabular}{ l  c }
        \hline
       & $m_{\PQt}$ [\GeV] \\ \hline
NNPDF3.0   &   $173.8^{+1.7}_{-1.8}$   \\
MMHT2014  &   $174.1^{+1.8}_{-2.0}$  \\
CT14      &   $174.3^{+2.1}_{-2.2}$ \\
\hline
\end{tabular}
 \label{tab:mtpolepdfs}
\end{table}

\section{Limits on top squark pair production}
\label{sec:stopsearch}

The SUSY models are predicated on the existence of partners for SM particles. A light top squark could contribute to the cancellation of
the quadratic divergences in the Higgs mass loop corrections~\cite{Nilles:1983ge}. SUSY scenarios with
a neutralino as LSP and a nearly mass-degenerate top squark provide one theoretically possible way to account for the observed relic abundance of dark matter~\cite{djoua,carena}. There are therefore strong motivations to search for a top squark with a mass close to, or even below, the \TeVns scale.

\begin{figure}[htbp]
\centering
\includegraphics[width=0.48\textwidth]{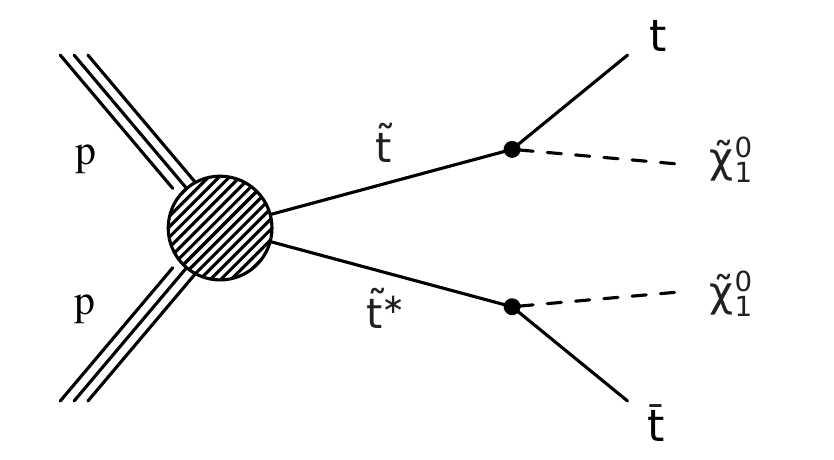}
\caption{Diagram displaying the top squark pair production at the LHC in the decay mode where each top squark decays to a top quark and a neutralino $\PSGczDo$.\label{fig:T2tt}}
\end{figure}

In the following, a SUSY model with $R$-parity conservation is considered, where top squarks are pair-produced via the strong interaction. The top squark decays into a top quark and the LSP, considered here as the lightest neutralino $\PSGczDo$. A simplified model is used, where the parameters are the top squark and neutralino masses~\cite{Alwall:2008ag,Alves:2011wf}.
The branching fraction of top squark into a top quark and a neutralino is assumed to be 100\%, and the top quark polarisation is assumed to be fully right-handed. A diagram of the process is shown in Fig.~\ref{fig:T2tt}.

Top squark pair production with the top squarks decaying into a top quark and
a neutralino could produce final states very similar to the one from
\ttbar production but with additional missing transverse
energy. If the difference between the masses of the top squark and
the neutralino is close to the top quark mass, the events would
have similar topologies to the SM \ttbar events. In such situations, direct top squark searches have low sensitivity because of the overwhelming \ttbar background.
However, from a very precise \ttbar cross section measurement, top squark pair events can be searched for by looking for a small excess in the measured
cross section compared to the SM expectation. The study presented here is complementary to the direct searches
performed by CMS~\cite{Chatrchyan:2013xna,Khachatryan:2015wza,Khachatryan:2016pup} and ATLAS~\cite{Aad:2014kra,Aad:2014bva,Aad:2014qaa}, as it is more sensitive in a mass region, $m(\PSQt) \approx m(\PSGczDo) + m_{\PQt}$, that is not accessible to conventional SUSY searches. Previous indirect searches in this mass region have been performed by the ATLAS collaboration~\cite{Aad:2014mfk,Aad:2015pfx}.

The 8\TeV data, analysed with the counting method (Section~\ref{sec:xsecext_evc}), are used to derive upper limits on the production cross section for the top squark pair production for different top squark masses.
The number of observed events in data is compared to the sum of SM \ttbar and background events and the expected yields from top squark pair production.

Top squark pair events generated with \MADGRAPH with up to two associated partons
are used for this study. The detector response is described using a fast simulation~\cite{Abdullin:2011zz}. In order to account for differences with the full simulation of the CMS detector used for all other samples, a correction for the b tagging SFs is applied. Furthermore, a 10\% uncertainty on the signal yields is added to account for the differences in lepton and trigger efficiencies between the fast and the full simulations. The signal samples are normalized according to the cross sections calculated at NLO+next-to-leading-logarithmic accuracy~\cite{NLO_Beenakker:1996ch,NLO_Kulesza,NLO_Kulesza:2009kq,NLO_Beenakker:2009ha,NLO_Beenakker:2011fu}.

The 95\% exclusion limits are calculated from Bayesian and modified
CL$_\mathrm{s}$ techniques implemented in the \textsc{Theta} framework~\cite{theta}.
The yields of events given in Table~\ref{tab:yieldsChannels} (where \ttbar MC events are normalised to the predicted NNLO cross section~\cite{mitov,topplusplus}) are used, accounting for all the systematic uncertainties described in Section~\ref{sec:syst}. The uncertainty of 3.5\% in
the theoretical \ttbar cross section is included to account for effects from
renormalisation and factorisation scale and PDF uncertainties in the
 calculation~\cite{mitov}.

\begin{figure}[htbp]
\centering
\includegraphics[width=0.48\textwidth]{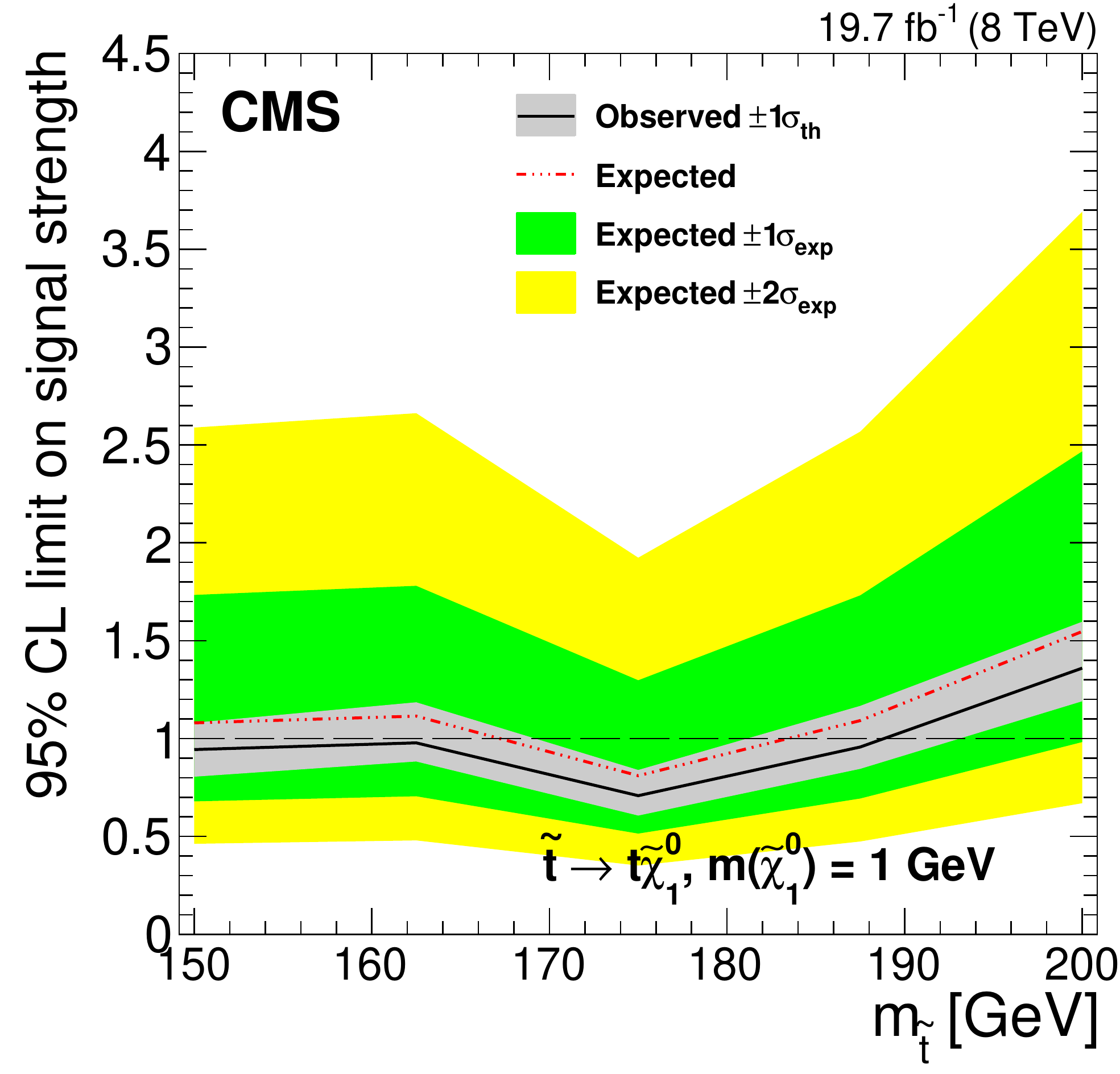}
\includegraphics[width=0.48\textwidth]{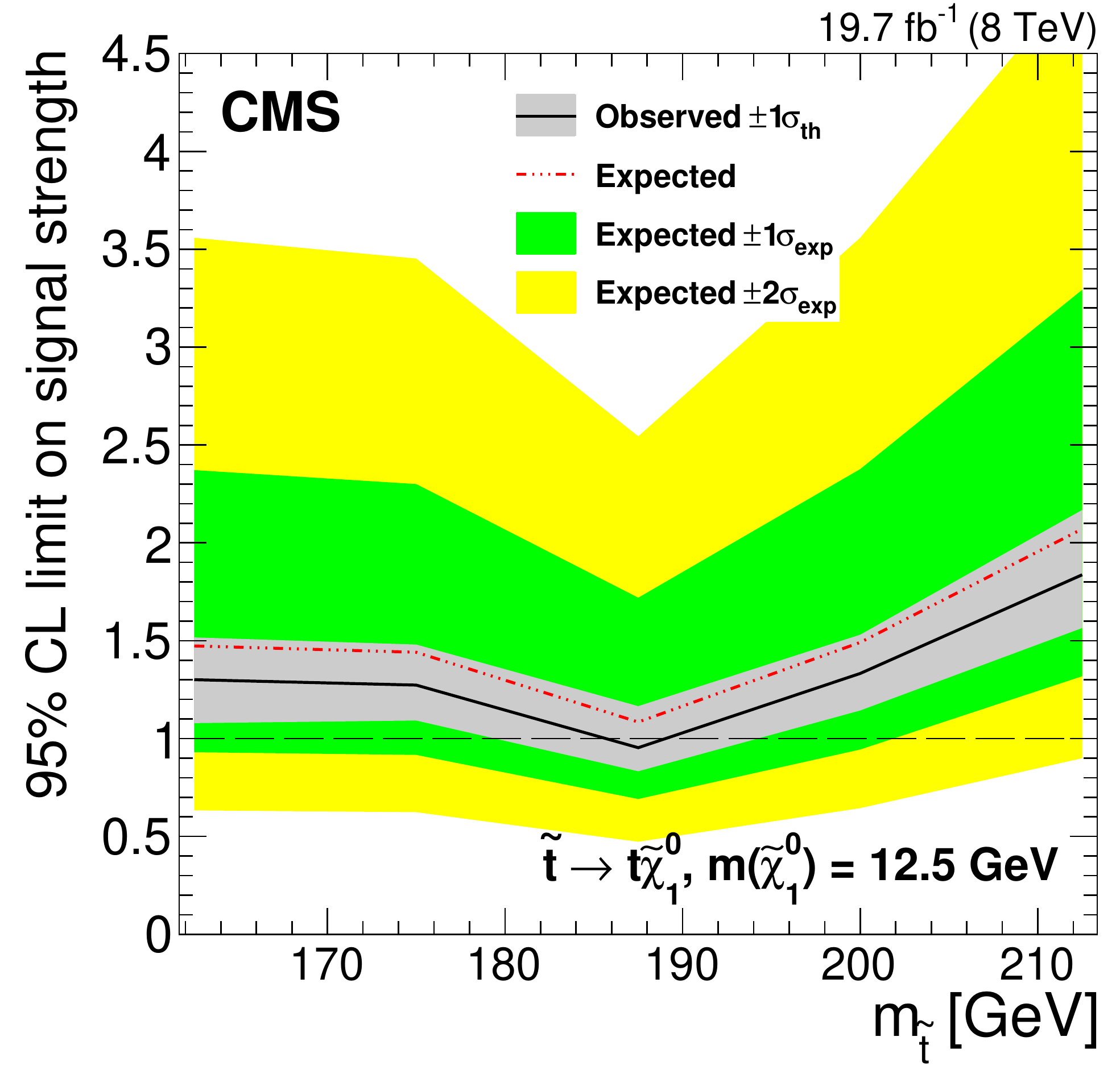}
\caption{Expected and observed limits at 95\% CL on the signal strength (see text) as a function of the top squark mass for neutralino masses of 1\GeV (left) and 12.5\GeV (right). The widest bands show the 68\% and 95\% CL ranges of the expected limit. The narrowest band quantifies the impact of the theoretical uncertainty in the cross section of the SUSY signal on the observed limit.\label{fig:stopLimits}}
\end{figure}

The observed and expected limits on the mass of the top squark for neutralino masses of 1 and 12.5~GeV are shown in
Fig.~\ref{fig:stopLimits}. The signal strength $\mu$ is defined as the ratio between the excluded cross section and the predicted one. Top squarks with masses below 189\GeV are excluded at 95\% CL for the neutralino mass of 1\GeV, and in the range 185--189\GeV for the neutralino mass of 12.5\GeV.

The effect of the top quark polarisation on the final result is studied by calculating the exclusion limits assuming that the top quarks are 100\% left-handed polarised. No significant differences are observed compared to the case of right-handed polarised top quarks.

\section{Summary}
\label{sec:conclusions}
A measurement of the inclusive
\ttbar production cross section in proton-proton
collisions at the LHC is presented using the full 2011--2012
data samples of 5.0\fbinv at  $\sqrt{s} = 7\TeV$
and 19.7\fbinv at  $\sqrt{s} = 8\TeV$. The analysis is performed in the
\emu
 channel using an improved cross section extraction method. The cross sections are determined with a binned likelihood
fit to the \pt distribution of the non-b-tagged jet with the lowest \pt among the selected jets in the event, using categories of number of b-tagged and additional non-b-tagged jets. Assuming a top quark mass of $172.5\GeV$, the results are
\begin{equation*}
\begin{aligned}
\sigma_{\ttbar} & = 
 173.6 \pm 2.1\stat ^{+\,4.5}_{-\,4.0} \syst\pm 3.8\lum\unit{pb},
\quad
\text{at $\sqrt{s}=7\TeV$ and}
\\
\sigma_{\ttbar} & = 
 244.9 \pm 1.4\stat ^{+\,6.3}_{-\,5.5}\syst\pm 6.4\lum\unit{pb},
\quad \text{at  $\sqrt{s}=8\TeV$,}
\end{aligned}
\end{equation*}
in good agreement with recent NNLO QCD calculations. The ratio of the cross sections at the two different values of $\sqrt{s}$\/ is determined to be $1.41 \pm 0.06$.
Moreover, the cross sections are measured in fiducial ranges defined by the transverse momentum and pseudorapidity requirements on the two charged leptons in the final state. The measurements constitute the most precise CMS results of \stt\
so far, and are competitive with recent ATLAS results~\cite{Aad:2014kva}.

The inclusive cross sections at 7 and 8\TeV
are used to determine the top quark pole mass via the dependence
of the theoretically predicted cross section on the mass, employing three different PDF sets.
The values of the mass are consistent between the three sets.
The most precise result, $173.8^{+1.7}_{-1.8}\GeV$, is obtained using the NNPDF3.0 PDF set.

The 8\TeV data are also used to
constrain the cross section of pair production of supersymmetric top squarks
with masses close to the top quark mass. No excess of event yields
with respect to the SM prediction is found,
and exclusion limits are presented as a function of the top squark mass for two different neutralino masses.

\begin{acknowledgments}
\hyphenation{Bundes-ministerium Forschungs-gemeinschaft Forschungs-zentren} We congratulate our colleagues in the CERN accelerator departments for the excellent performance of the LHC and thank the technical and administrative staffs at CERN and at other CMS institutes for their contributions to the success of the CMS effort. In addition, we gratefully acknowledge the computing centres and personnel of the Worldwide LHC Computing Grid for delivering so effectively the computing infrastructure essential to our analyses. Finally, we acknowledge the enduring support for the construction and operation of the LHC and the CMS detector provided by the following funding agencies: the Austrian Federal Ministry of Science, Research and Economy and the Austrian Science Fund; the Belgian Fonds de la Recherche Scientifique, and Fonds voor Wetenschappelijk Onderzoek; the Brazilian Funding Agencies (CNPq, CAPES, FAPERJ, and FAPESP); the Bulgarian Ministry of Education and Science; CERN; the Chinese Academy of Sciences, Ministry of Science and Technology, and National Natural Science Foundation of China; the Colombian Funding Agency (COLCIENCIAS); the Croatian Ministry of Science, Education and Sport, and the Croatian Science Foundation; the Research Promotion Foundation, Cyprus; the Ministry of Education and Research, Estonian Research Council via IUT23-4 and IUT23-6 and European Regional Development Fund, Estonia; the Academy of Finland, Finnish Ministry of Education and Culture, and Helsinki Institute of Physics; the Institut National de Physique Nucl\'eaire et de Physique des Particules~/~CNRS, and Commissariat \`a l'\'Energie Atomique et aux \'Energies Alternatives~/~CEA, France; the Bundesministerium f\"ur Bildung und Forschung, Deutsche Forschungsgemeinschaft, and Helmholtz-Gemeinschaft Deutscher Forschungszentren, Germany; the General Secretariat for Research and Technology, Greece; the National Scientific Research Foundation, and National Innovation Office, Hungary; the Department of Atomic Energy and the Department of Science and Technology, India; the Institute for Studies in Theoretical Physics and Mathematics, Iran; the Science Foundation, Ireland; the Istituto Nazionale di Fisica Nucleare, Italy; the Ministry of Science, ICT and Future Planning, and National Research Foundation (NRF), Republic of Korea; the Lithuanian Academy of Sciences; the Ministry of Education, and University of Malaya (Malaysia); the Mexican Funding Agencies (CINVESTAV, CONACYT, SEP, and UASLP-FAI); the Ministry of Business, Innovation and Employment, New Zealand; the Pakistan Atomic Energy Commission; the Ministry of Science and Higher Education and the National Science Centre, Poland; the Funda\c{c}\~ao para a Ci\^encia e a Tecnologia, Portugal; JINR, Dubna; the Ministry of Education and Science of the Russian Federation, the Federal Agency of Atomic Energy of the Russian Federation, Russian Academy of Sciences, and the Russian Foundation for Basic Research; the Ministry of Education, Science and Technological Development of Serbia; the Secretar\'{\i}a de Estado de Investigaci\'on, Desarrollo e Innovaci\'on and Programa Consolider-Ingenio 2010, Spain; the Swiss Funding Agencies (ETH Board, ETH Zurich, PSI, SNF, UniZH, Canton Zurich, and SER); the Ministry of Science and Technology, Taipei; the Thailand Center of Excellence in Physics, the Institute for the Promotion of Teaching Science and Technology of Thailand, Special Task Force for Activating Research and the National Science and Technology Development Agency of Thailand; the Scientific and Technical Research Council of Turkey, and Turkish Atomic Energy Authority; the National Academy of Sciences of Ukraine, and State Fund for Fundamental Researches, Ukraine; the Science and Technology Facilities Council, UK; the US Department of Energy, and the US National Science Foundation.

Individuals have received support from the Marie-Curie programme and the European Research Council and EPLANET (European Union); the Leventis Foundation; the A. P. Sloan Foundation; the Alexander von Humboldt Foundation; the Belgian Federal Science Policy Office; the Fonds pour la Formation \`a la Recherche dans l'Industrie et dans l'Agriculture (FRIA-Belgium); the Agentschap voor Innovatie door Wetenschap en Technologie (IWT-Belgium); the Ministry of Education, Youth and Sports (MEYS) of the Czech Republic; the Council of Science and Industrial Research, India; the HOMING PLUS programme of the Foundation for Polish Science, cofinanced from European Union, Regional Development Fund; the OPUS programme of the National Science Center (Poland); the Compagnia di San Paolo (Torino); MIUR project 20108T4XTM (Italy); the Thalis and Aristeia programmes cofinanced by EU-ESF and the Greek NSRF; the National Priorities Research Program by Qatar National Research Fund; the Programa Clar\'in-COFUND del Principado de Asturias; the Rachadapisek Sompot Fund for Postdoctoral Fellowship, Chulalongkorn University (Thailand); the Chulalongkorn Academic into Its 2nd Century Project Advancement Project (Thailand); and the Welch Foundation, contract C-1845.
\end{acknowledgments}
\bibliography{auto_generated}

\cleardoublepage \appendix\section{The CMS Collaboration \label{app:collab}}\begin{sloppypar}\hyphenpenalty=5000\widowpenalty=500\clubpenalty=5000\textbf{Yerevan Physics Institute,  Yerevan,  Armenia}\\*[0pt]
V.~Khachatryan, A.M.~Sirunyan, A.~Tumasyan
\vskip\cmsinstskip
\textbf{Institut f\"{u}r Hochenergiephysik der OeAW,  Wien,  Austria}\\*[0pt]
W.~Adam, E.~Asilar, T.~Bergauer, J.~Brandstetter, E.~Brondolin, M.~Dragicevic, J.~Er\"{o}, M.~Flechl, M.~Friedl, R.~Fr\"{u}hwirth\cmsAuthorMark{1}, V.M.~Ghete, C.~Hartl, N.~H\"{o}rmann, J.~Hrubec, M.~Jeitler\cmsAuthorMark{1}, A.~K\"{o}nig, M.~Krammer\cmsAuthorMark{1}, I.~Kr\"{a}tschmer, D.~Liko, T.~Matsushita, I.~Mikulec, D.~Rabady, N.~Rad, B.~Rahbaran, H.~Rohringer, J.~Schieck\cmsAuthorMark{1}, J.~Strauss, W.~Treberer-Treberspurg, W.~Waltenberger, C.-E.~Wulz\cmsAuthorMark{1}
% ...
\vskip\cmsinstskip
% ...
1:~~Also at Vienna University of Technology, Vienna, Austria\\
% ...

\end{sloppypar}
\end{document}